\documentclass[aps,prb,twocolumn,groupedaddress,floatfix]{revtex4-1}

\usepackage{amsmath,amssymb}
\usepackage{graphicx}
\usepackage{color}
\usepackage{hyperref}

\begin{document}


\title{Generation of Schr\"odinger's cat states in a planar semiconductor heterostructure}


\author{J. Paw\l{}owski}
\author{M. G\'orski}
\author{G. Skowron}
\author{S. Bednarek}
\affiliation{
	Faculty of Physics and Applied Computer Science,
	AGH University of Science and Technology, Krak\'{o}w, Poland}



\date{\today}

\begin{abstract}
We propose a nanodevice based on a typical planar semiconductor heterostructure with lateral confinement potential created by voltages applied to local gates. We show how to obtain near parabolical confinement along the nanodevice, and how to use coherent states of the harmonic oscillator for spatial separation of electron densities corresponding to opposite spin directions. In such a way, an entangled state of Schr\"odinger's cat type is created. We have performed simulations of a realistic nanodevice model by numerical solving the time-dependent Schr\"odinger's equation together with simultaneous tracking of the controllable confinement potential via solution of the Poisson's equation at every time step. 
\end{abstract}


\maketitle

\section{Introduction}
Control and manipulation of single electrons trapped in semiconductor nanostructures attract
much attention due to potential applications in spintronics \cite{hanson, zutic} or quantum
computing \cite{ladd}. This also enables us to examine numerous fundamental physical phenomena and
discover a new physics, e.g. topological effects \cite{hasan, nori} or exotic quasi-particles
\cite{cz1, cz2, cz3, cz4}.

An electron qubit can be represented in many different ways \cite{owen2}. In the case of a charge
qubit \cite{chq1, chq2, chq3}, two basis states are defined as the presence of an electron on one of two sides of
a double quantum dot or wire structure. However, it is more likely that the spin degree of freedom will be used as
a bit carrier in the quantum computer, i.e. spin qubit\cite{loss, loss1}. This explains the necessity for
precise electron control to perform operations on such qubits, or to couple them into registers
\cite{coupled, floating, cp2, cp3, cp4, cp5, bramki1}, and transmit information between individual
registers of the quantum computer \cite{metoda, bramki2, praceprof3, praceprof4}. Spin-orbit interaction (SOI) of Rashba type
(RSOI) \cite{rashba1, bychkov,rashba2, winkler}, which couples the orbital and spin degrees of freedom of an electron,
allows for effective manipulation of a spin qubit \cite{nowack, np0, np1, japarize, lopez, sherman, ramsak, szaszko, pawlowski1, pawlowski2}.

An important issue in quantum computing is the generation of entangled states. The Schr\"odinger's cat state is a notable example \cite{winelandprzeglad, kot1, ralph, liao}.
This has been successfully generated in quantum optics using coherent states \cite{winelandzrodlo, blatt, wineland1, sackett, jeong}. 
The possibility to create coherent states has also been examined in solid-state systems \cite{owen1, owen2, pysh, liao, kat, laskotek}. 
However, creation of their combination with opposite spin, namely the Schr\"odinger's cat state, poses a
great challenge due to its high instability \cite{winelandprzeglad, blatt1, wineland1, inni}.

In this paper, we propose to repeat these quantum optics experiments in a solid state system. We show the possibility for the
creation of a Schr\"odinger's cat state in a typical and easily obtainable semiconductor
heterostructure. 
This is an extension to the method introduced in [\onlinecite{laskotek}], here developed on heterostructures, which are much more easily scalable.
Control over an electron is achieved all electrically by
applying voltages to local gates. The spin separation of coherent states,
forming in superposition the Schr\"odinger's cat state, is achieved in
the nanostructure with an electrically controlled Rashba spin-orbit coupling.

The paper is organized as follows. In the first part (Sec.~\ref{firstpart}) of the article we employ a one-dimensional (1D) approximation with modeled
potentials and the method for the generation of the Schr\"odinger's cat state is discussed only
qualitatively. In the latter part (Sec.~\ref{secondpart}) we propose a nanodevice based on a planar
semiconductor heterostructure. The design includes geometrical details and the realistic values for parameters of the materials used.
Potentials are calculated using the Poisson's equation with all important effects included. The results from this part are quantitative in nature.

\section{Explaining the effect with a simplified model\label{firstpart}}

\subsection{Simplified one-dimensional model}
The Hamiltonian of a single electron trapped in a quantum wire---a 1D structure---oriented along the $x$-axis has the following form:
\begin{equation}\label{schrod1}
 H(x,t) =\left( -\frac{\hbar^2}{2m}\frac{\partial^2}{\partial x^2}+V(x)\right)\!1_2 +H_{\mathrm{so}}(t),
\end{equation}
where $V(x)$ constitutes the potential energy of an additional confinement along the wire and $H_{\mathrm{so}}(t)$ describes the spin-orbit interaction. If the quantum
wire axis is oriented in the $[111]$ crystallographic direction, the Dresselhaus interaction becomes
negligible \cite{np1, winkler}. Now if we apply an electric field along the $y$-axis, an RSOI is
introduced. It can be described by the following Hamiltonian:
\begin{equation}\label{hso}
H_{\mathrm{\mathrm{so}}}(t) = -\frac{\alpha_{\mathrm{\mathrm{so}}}|e|}{\hbar}E_y(t)\,p_x\sigma_z,
\end{equation}
with the RSOI material coefficient $\alpha_{\mathrm{\mathrm{so}}}$, the electric field $E_y(t)$ perpendicular to the wire direction, and the electron momentum operator $p_x$ together with the the Pauli $z$-matrix $\sigma_z$.

If the confinement
potential along the wire has a parabolic shape $V(x)=m\omega^2x^2/2$, we can solve the
eigenequation of the Hamiltonian (\ref{schrod1}) analytically in the momentum representation and then return to the
position representation\cite{laskotek}. The energy of the ground state is doubly degenerated with respect to spin values.
The ground state wavefunction in the position representation takes the form of a Gaussian multiplied
by a plane wave. Depending on the spin $z$-projection its wavenumber is either positive or
negative. The two basis functions corresponding to the ground state have the two-row spinor form
\begin{align}
\Psi_{\uparrow}(x)&=\left(\frac{2\beta}{\pi}\right)^\frac{1}{4}\begin{pmatrix}1\\0\end{pmatrix}e^{-\beta x^2}e^{iqx}, \label{groundstate1}\\
\Psi_{\downarrow}(x)&=\left(\frac{2\beta}{\pi}\right)^\frac{1}{4}\begin{pmatrix}0\\1\end{pmatrix}e^{-\beta x^2}e^{-iqx},\label{groundstate2}
\end{align}
with $\beta=\frac{m\omega}{2\hbar}$ and the wavenumber $q=\frac{m\alpha_{\mathrm{\mathrm{so}}}|e|E_y(t=0)}{\hbar^2}$.

We should note that the wavenumber $q$ does not depend on the harmonic potential curvature (frequency $\omega/2\pi$). 
Indeed, for any confinement potential $V(x)$ the eigenstate of (\ref{schrod1}) has the general form $\varphi(x) e^{\pm iqx}$, where $\varphi(x)$ is an eigenfunction calculated without the spin-orbit interaction. SOI introduces a spin-dependent \textit{displacement} in the momentum (space).
Due to the degeneration, any linear combination of both basis states corresponds to the same energy. The SOI merely introduces an energy correction $\Delta \mathcal{E}=-\frac{\hbar^2q^2}{2m}$.

\subsection{Semiconductor electron soliton -- an inducton}
\begin{figure}
\centering
\includegraphics[width=0.45\textwidth]{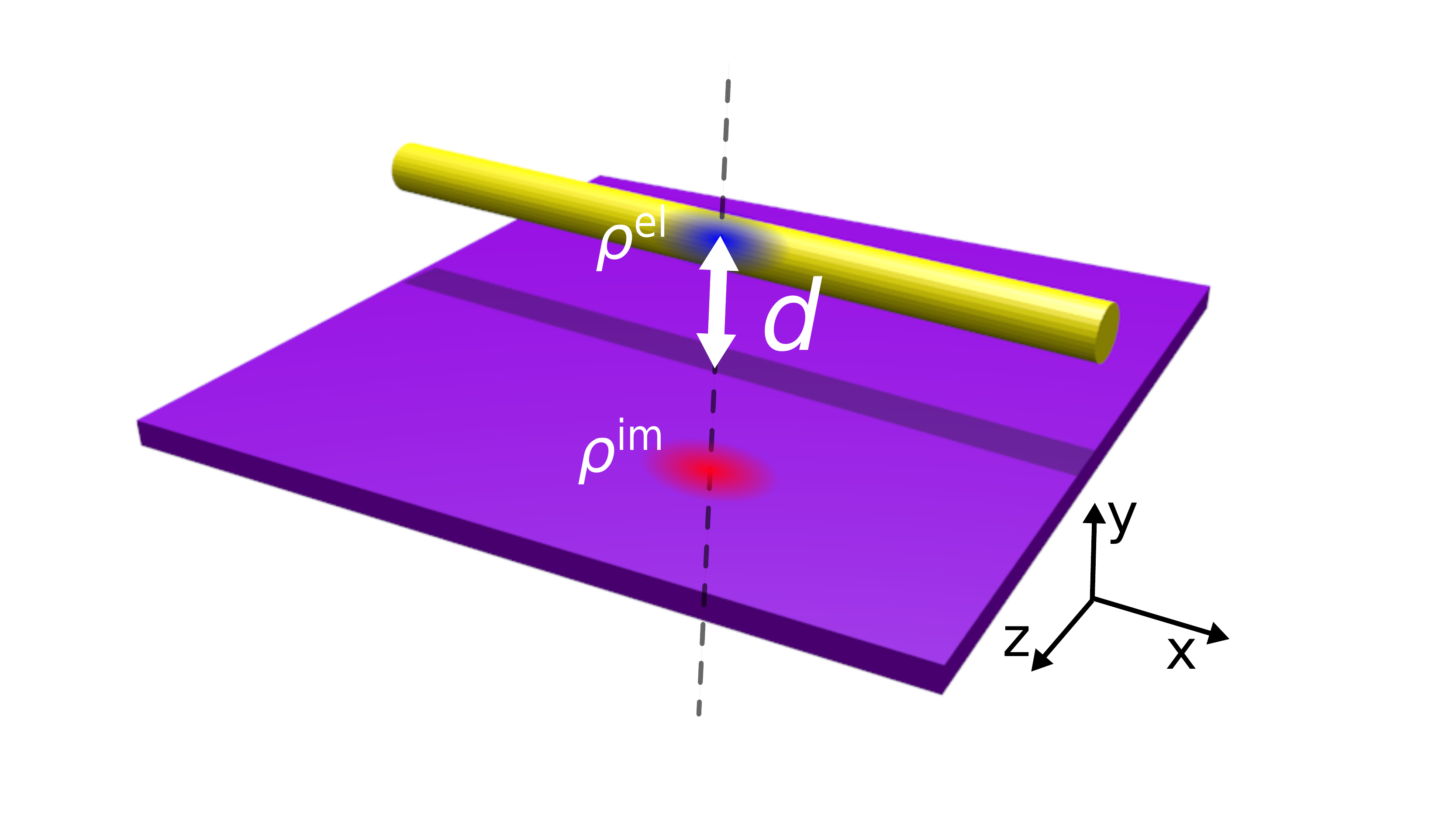}
\caption{\label{fig18}Scheme depicting the simplified 1D model with a quantum wire above a conducting plate at distance $d$.}
\end{figure}
We assume that the quantum wire is aligned in parallel to a metallic gate, as shown in Fig.~\ref{fig18}. If we trap an
electron inside the wire, an opposite charge will be induced on the surface of the conductor.
This effect can be described within the mean field self-consistent approximation\cite{indukton1}.
The induced charge attracts the electron and causes the mean electric field to have a component directed
toward the center of the electron density, resulting in a self-focusing of the electron
wavefunction. Thus, an electron soliton called an inducton\cite{indukton1,indukton} is formed. The electron becomes trapped under the gate, creating a
stable Gaussian-like wavepacket of finite size, capable of moving without
changing its shape. 
Because of such localization we can relocate the electron
within the wire in a controllable manner \cite{metoda, bramki2, bramki3}. 
Moreover, inductons and their spins can be used as quantum bit carriers in the quantum computer \cite{indukton, bramki1, bramki2, bramki3}. 

This mean field, in which the electron is located, is generated by the electron (charge) itself. Thus, the field is calculated in a self-consistent way.
If the gate is an infinite
conducting plate, the potential created by the induced charge can be described using the
image charge method. The induced charge is replaced by an image charge which is a
reflection (against the surface of the conductor) of the primary charge density from the
quantum wire \cite{indukton1}. For a quantum wire placed at distance $d$ from
the conducting plate, the potential energy created by interaction with the image charge can be
expressed in the following way:
\begin{equation}
U^{\mathrm{ind}}(x)=\frac{-|e|}{4\pi\epsilon\epsilon_0}\int\frac{\rho^\mathrm{im}(x^\prime)}{\sqrt{(x-x^\prime)^2+4d^2}}\,\mathrm{d}x^\prime,
\label{im}
\end{equation}
with the image charge density $\rho^\mathrm{im}(x)$ being a mirrored version of the electron charge from the quantum wire $\rho^\mathrm{el}(x)$---see Fig.~\ref{fig18}, and calculated as
\begin{equation}
\rho^\mathrm{im}(x)=-\rho^\mathrm{el}(x)=|e||\Psi(x)|^2.
\label{imm}
\end{equation}
If there is no additional external potential in (Eq.~\ref{schrod1}), only the interaction energy with the image charge is present: $V(x)=U^\mathrm{ind}(x)$. Since the image charge method is applicable only for cases of an infinite interface between a dielectric medium and a metallic plate, this approach is used only for model potentials in the first part of this article (Sec. \ref{firstpart}). For actual realizations of the nanodevice (Sec. \ref{secondpart}), the potential inside the quantum wire is exactly calculated using the Poisson's equation. This method is slightly more complex; however, it guarantees correct inclusion of the induced charge on gates of any shape or dimension \cite{zgodnosc}. 

The potential energy for the electron ground state, originating from the induced charge present on the gate parallel to the quantum wire, takes an approximately parabolic shape \cite{indukton1}. Therefore, if this potential energy, and the SOI generated by the electric field are taken into account, the wavefunction of the electron ground state in the wire is a Gaussian multiplied by a plane wave as in (Eqs.~\ref{groundstate1} or \ref{groundstate2}).
As noted, we get such multiplied eigenfunctions  $\varphi(x)e^{\pm iqx}$ for any form of the confining potential, also for the induced one $U^\mathrm{ind} (x,|\varphi|^2)$.

\subsection{Control of the electron motion}
Let us assume that the electron spin is directed along the $z$-axis (spin $z$-projection $s_z=\hbar/2$) with a nonzero electric field $E_y$. In such a case the ground state wavefunction assumes the form of (Eq. \ref{groundstate1}). This is a stationary state and the wavepacket remains fixed even though the wavefunction is a Gaussian multiplied by a plane wave. This happens because motion is blocked by the SOI included in the Hamiltonian.
However, if we disable the electric field \textit{abruptly} (in a non-adiabatic manner), inserting $E_y=0$, the SOI disappears and the electron starts moving in the $x$-direction.
Now the wavefunction is a gaussian multiplied by a plane wave. Such a wave packet travels at a constant speed $v=\frac{\hbar q}{m}$:
\begin{align}
&\Psi_\uparrow(x,t)=\left(\frac{2\beta}{\pi}\right)^{\frac{1}{4}}\!\!\begin{pmatrix}1\\0\end{pmatrix}\! e^{-\beta (x-vt)^2}e^{iqx}e^{-\frac{i\mathcal{E}t}{\hbar}}= \nonumber \\
&=\Psi_\uparrow(x-vt,0)\,e^{\frac{i\mathcal{E}t}{\hbar}}, \label{galileo}
\end{align}
with $\mathcal{E}=\frac{\hbar^2 q^2}{2m}$.
Let us note that after inserting this time-dependent wavefunction (\ref{galileo}) into the expression (\ref{imm}), the image charge density moves at the same speed $\rho^\mathrm{im}(x,t)=\rho^\mathrm{im} (x-vt,0)$. Inserting this into (Eq.~\ref{im}) gives a time-dependent potential obeying a similar relation
\begin{align*}
V(x,t)={}&U^\mathrm{ind}(x,|\Psi_\uparrow(x,t)|^2)=\\
={}&U^\mathrm{ind}(x,|\Psi_\uparrow(x-vt,0)|^2)=V(x-vt,0). 
\end{align*}
Thus the induced potential follows the wave packet at the same speed. This allows both the wave packet and the potential for moving in space without changing shape. In other words, the nonlinear Schr\"odinger equation with the potential $U^\mathrm{ind}(x,|\psi(x,t)|^2)$ is Galilean invariant in sense, that if $\psi(x,0)$ is a stationary solution, the time-dependent $\psi(x,t)=\psi(x-vt,0)\exp(iqx-i\mathcal{E}t/\hbar)$ is also its solution.

Setting $E_y$ back to the previous value stops the electron again, while using a greater value for $E_y$ forces movement in the opposite direction. Thus, we gain a method to control the electron motion using an electric field perpendicular to the direction of motion. 
The time evolution of the electron packet is obtained by numerically solving the time-dependent Schr\"{o}dinger equation $i\hbar\frac{\partial}{\partial t}\Psi(x,t)=H(x,t)\Psi(x,t)$, for the Hamiltonian (\ref{schrod1}) with variable spin-orbit part $H_{\mathrm{\mathrm{so}}}(t)$. The Schr\"{o}dinger equation is solved  self consistently with the induced $V(x,t)$ potential, which in turn depends on the electron density $|\Psi(x,t)|^2$. In the latter part (Sec.~\ref{secondpart}) this will be replaced by self consistency with the Poisson equation.

\begin{figure}[t]
\centering
\includegraphics[width=0.45\textwidth]{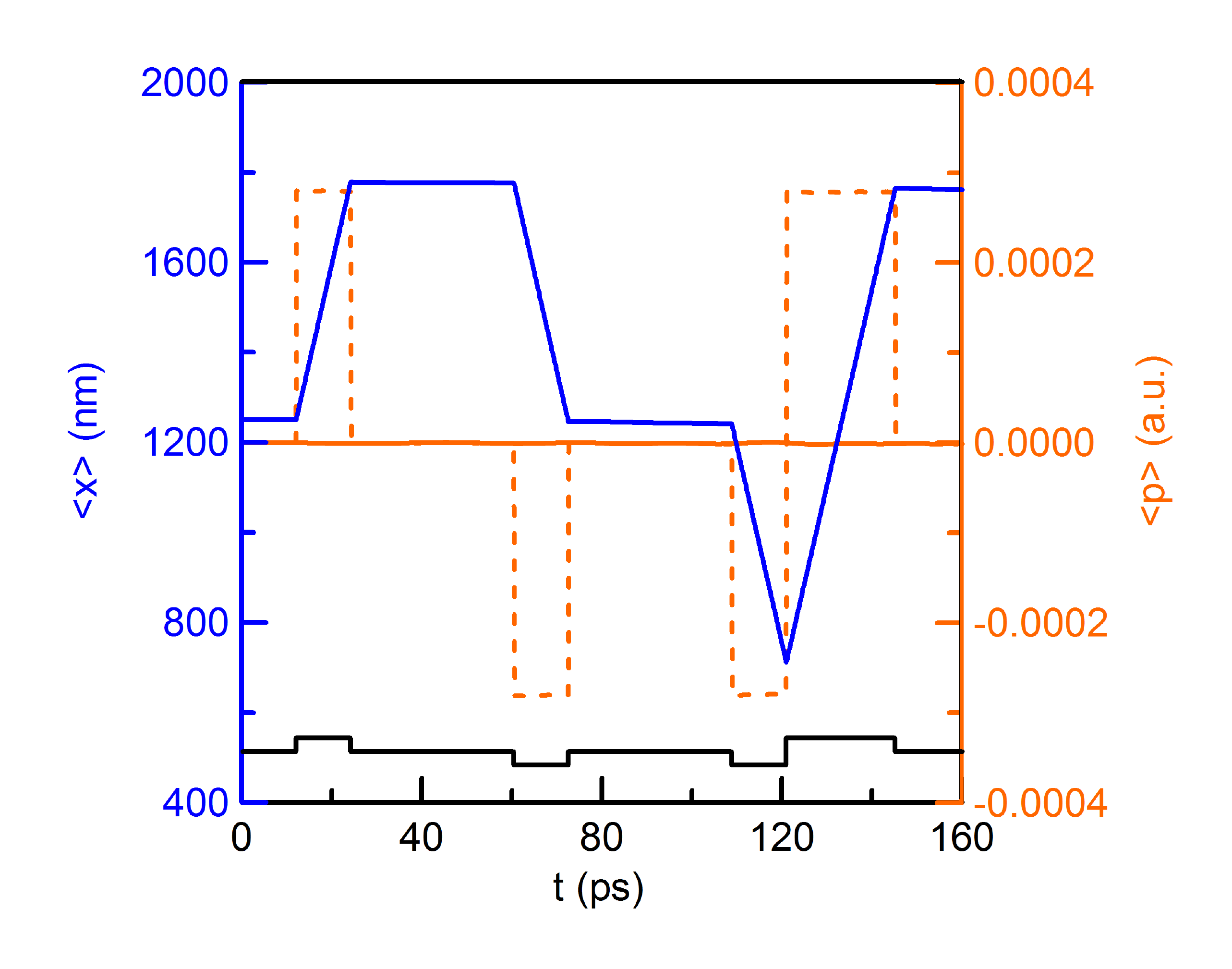}
\caption{Simulation of the electron motion induced by sudden changes of the electric field $E_y(t)$ perpendicular to the quantum wire denoted by black curve. The other curves denotes position of the electron $\langle x \rangle$ (blue curve), the expectation value the electron momentum $\langle p \rangle$ (solid red), and the classical momentum calculated as a time derivative of the position multiplied by the electron mass $m\frac{d}{dt}\langle x \rangle$ (dashed red).}
\label{fig1}
\end{figure}

Now we assume that the ground state is generated with $E_y=0$. This time the wavefunction is again a Gaussian but no longer multiplied by a plane wave ($q=0$). We propel the electron by setting non-zero $E_y$. Fig.~\ref{fig1} shows an electron motion induced by changes of the electric field. The black curve denotes the value of $E_y$, the blue curve denotes the expectation value of position $\langle x \rangle$, the solid red one denotes the expectation value of momentum $\langle p \rangle$ and the dashed red one denotes the classical momentum calculated as a time derivative of the position $\frac{d}{dt}\langle x \rangle$ multiplied by the electron mass $m$.
\begin{figure}[b]
\centering
\includegraphics[width=0.45\textwidth]{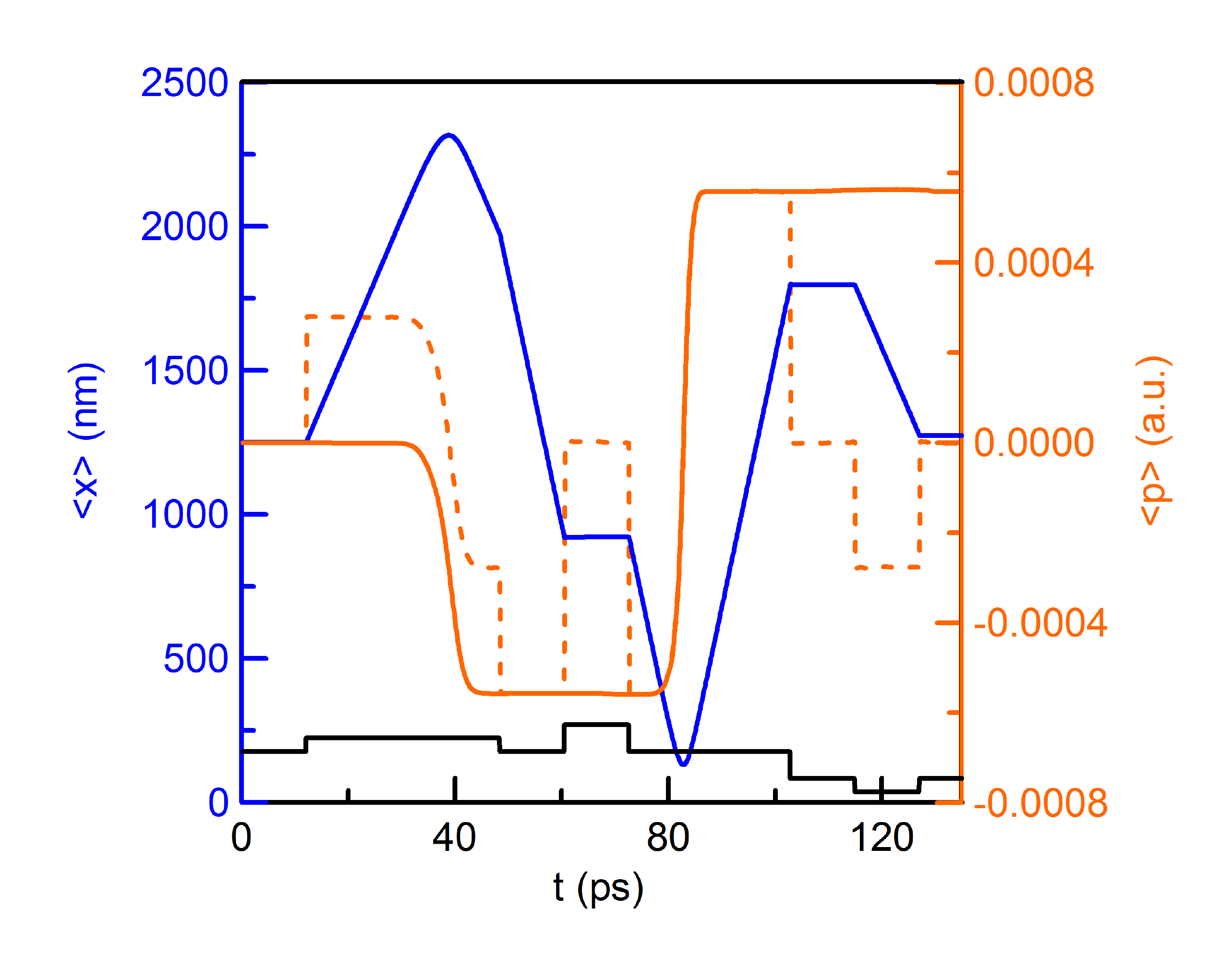}
\caption{\label{fig2}Simulation of the electron motion including reflection off the walls at the wire ends. Markings as in Fig.~\ref{fig1}.}
\end{figure}
Initially, we apply no electric field ($E_y=0$) and the electron remains stable. At $t=10$~ps we set a positive value for $E_y$, which sets the electron in motion in the direction of positive values of $x$. At $t=20$~ps, we set the electric field back to zero and the electron halts. Finally at $t=60$~ps, the electric field is set to $-E_y$, which induces movement toward negative values of $x$. Further manipulation of the electric field alters the velocity and the direction of the electron motion. We should notice that, despite the movement occurring at various velocities (corresponding to different values of the classical momentum $m\frac{d}{dt}\langle x \rangle$), the expectation value of momentum operator $\langle p \rangle$ remains zero, which means that the wavefunction is only a Gaussian, yet not multiplied by a plane wave ($q(t)=0$). However, if the electric field is non-zero (SOI present), this state is no longer stationary. Therefore, the electron motion is initiated as a result of the change of the Hamiltonian, and not the wavefunction.

The situation changes significantly if we allow the electron to reflect off the wall of the potential formed at the wire ends. 
In the Fig.~\ref{fig2}, we can track the motion of the electron in such a case with reflections occurring at $t=40$~ps and $t=83$~ps. After reflection, the electron moves with the same speed but in the opposite direction to the initial one. This indicates that the wavefunction has been effectively multiplied by a plane wave with a doubled wavenumber $2q$. The expectation value of the momentum operator $\langle p \rangle$ is no longer zero, but still inconsistent with the classical momentum $m\frac{d}{dt}\langle x \rangle$. At the moment $t=120$~ps, the directions of both these  quantities are actually \textit{opposite} to each other.

\subsection{Accelerating the electron -- a synchrotron}
\begin{figure}[b]
\centering
\includegraphics[width=0.45\textwidth]{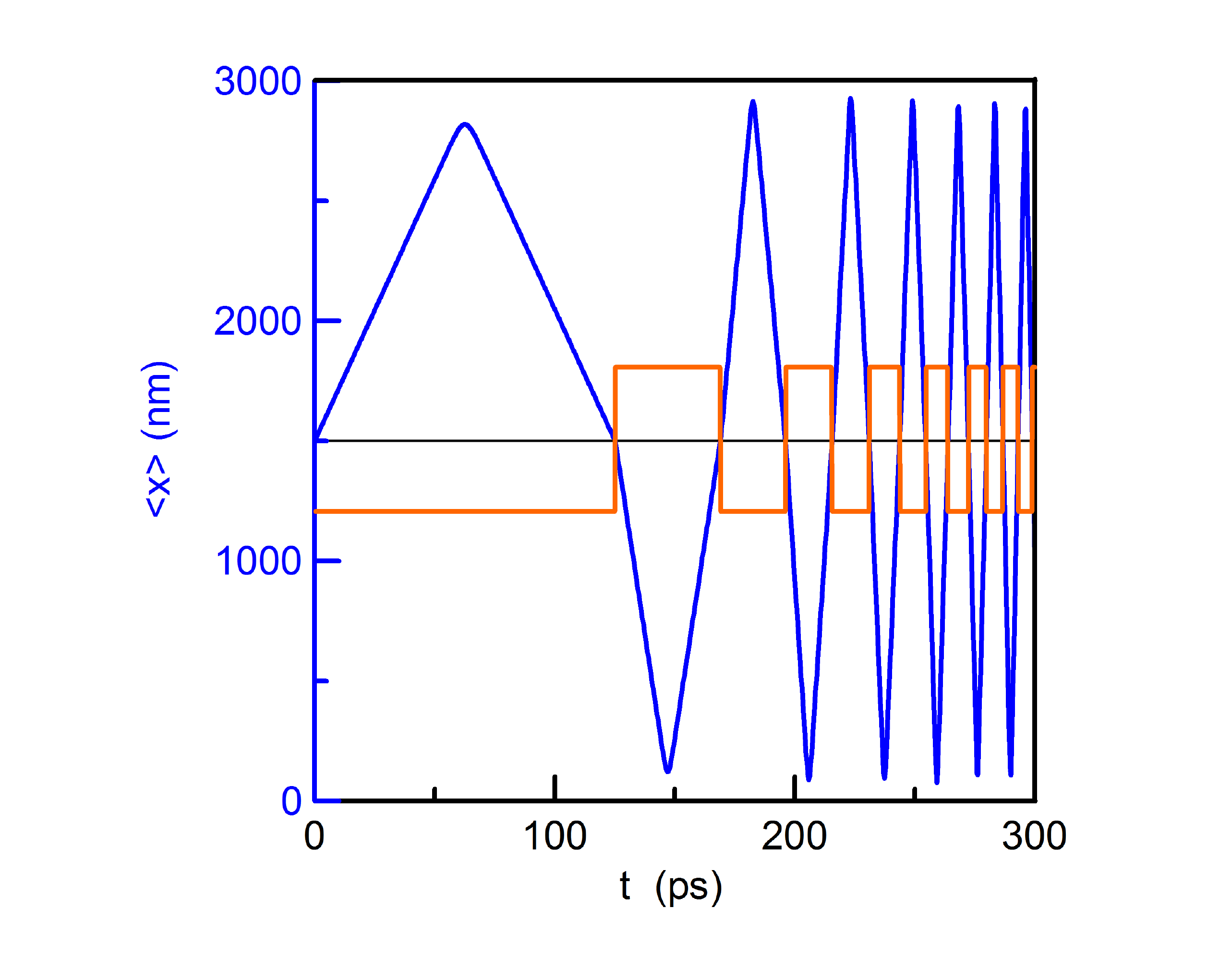}
\caption{\label{fig3}Acceleration of the electron by square pulses of the electric field $E_y$, denoted by the red curve; while the blue curve shows the expectation value of the electron wavepacket position $\langle x \rangle$.}
\end{figure}

The change of the wavefunction due to reflection can be exploited for wavepacket acceleration. In Fig.~\ref{fig3} we see a simulation of the electron motion induced in a rectangular potential well by square pulses of the electric field. The pulses have constant amplitude and duration carefully tuned to the moment of transition of the wavepacket through the central point of the wire. With every change of direction of the electric field, the electron is being accelerated.
With increasing speed of the electron, the time between reflections decreases; hence, the changes of the electric field must be performed at decreasing periods of time. 

\begin{figure}[b]
\centering
\includegraphics[width=0.45\textwidth]{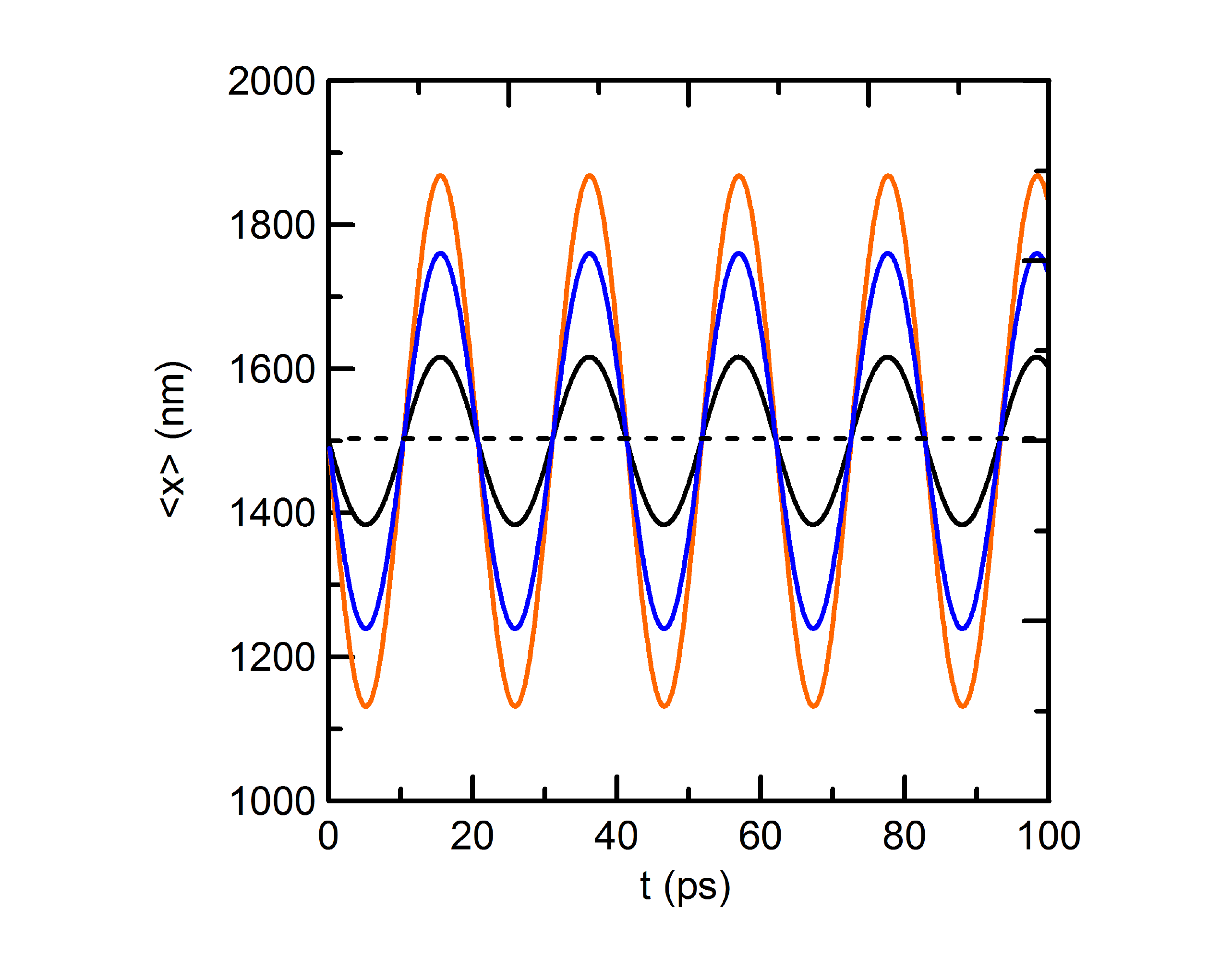}
\caption{\label{fig4}Motion of the electron in an external parabolic potential $U^\mathrm{ext}(x)=m\omega^2x^2/2$, calculated for three different values for the wavenumber $q$ in (Eq. \ref{parabwf2}).}
\end{figure}

This inconvenience can be mitigated by putting the electron in an external parabolic confinement potential $U^\mathrm{ext}(x)=m\omega^2x^2/2$, making $V(x)=U^\mathrm{ind}(x) + U^\mathrm{ext}(x)$. In this case, with no SOI, the ground state wavefunction of the electron assumes a Gaussian form:
\begin{equation}\label{parabwf}
\Psi(x)\equiv\langle x|p=0\rangle=\left(\frac{2\beta'}{\pi}\right)^\frac{1}{4}e^{-\beta' x^2}.
\end{equation}
Note that parameter $\beta'$ is renormalized by self-interaction, while eigenfrequency remains $\omega/2\pi$.
Multiplication of this Gaussian by a plane wave sets the electron in an oscillatory motion with an amplitude dependent on $q$, yielding
\begin{equation}\label{parabwf2}
\langle x|p=q\hbar\rangle=\left(\frac{2\beta'}{\pi}\right)^\frac{1}{4}e^{-\beta' x^2}e^{iqx}.
\end{equation}
Fig.~\ref{fig4} shows the motion of the electron initially set as the ground state of the harmonic oscillator, without SOI but multiplied by a plane wave with three different values for $q$.
Regardless of the actual value of the wavenumber $q$ (and energy) the periods of oscillations remain the same, as for a classical particle. This effect can be used for motion synchronization.

\begin{figure}[t]
\centering
\includegraphics[width=0.45\textwidth]{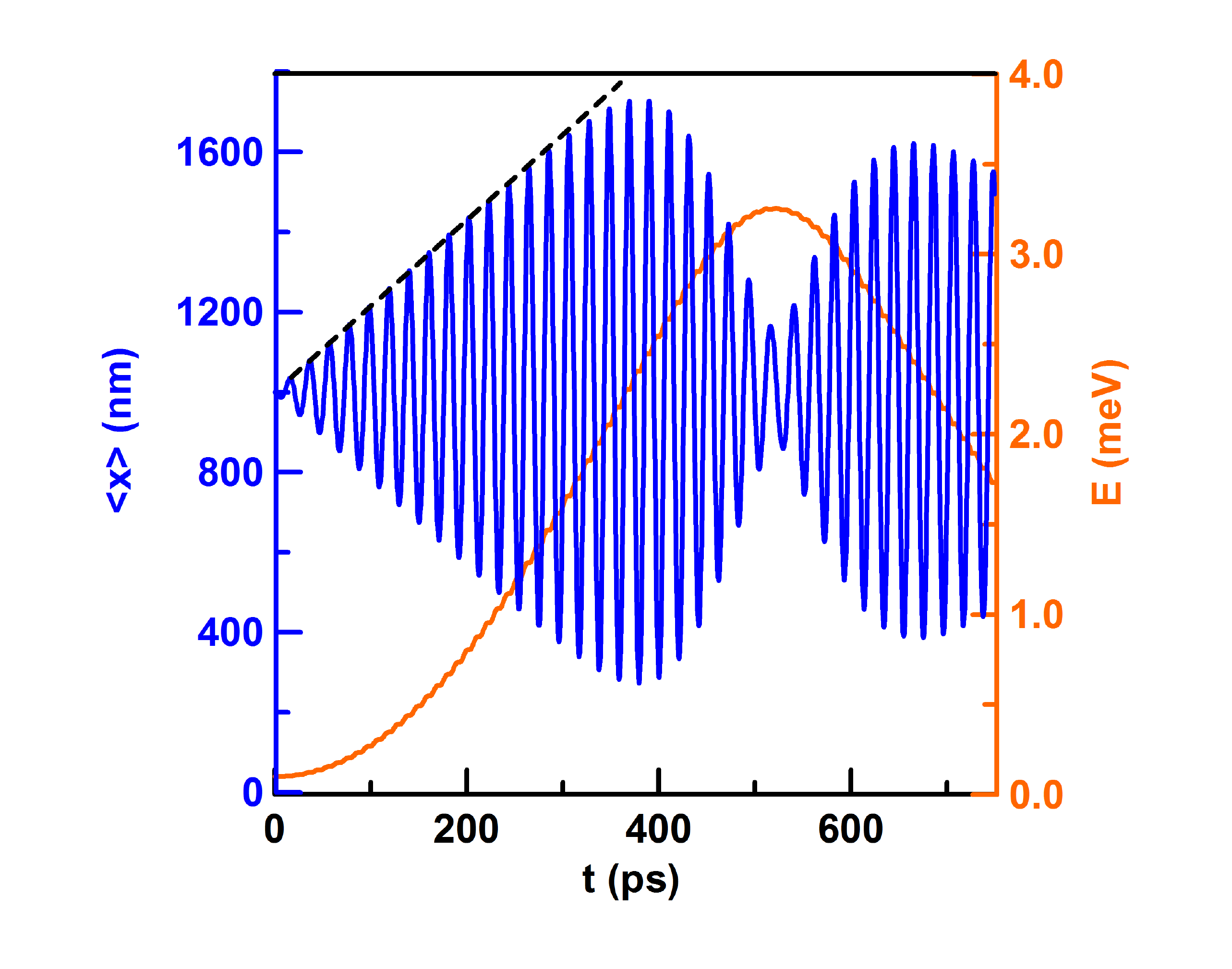}
\caption{\label{fig5}Acceleration of the electron confined in a harmonic potential $m\omega^2x^2/2$ using a sinusoidally varying electric field $E_y(t)$. The blue line denotes $\langle x \rangle$ and the red one---the electron energy.}
\end{figure}

If the electron is confined in a harmonic potential with additional spin-orbit coupling, varying sinusoidally with frequency $\omega/2\pi$ consistent with the frequency of the harmonic potential, we can accelerate the electron to high velocities using only low gate voltages (and thus we obtain a synchrotron-like device). This can be achieved using a sinusoidally oscillating electric field $E_y(t)=E_0\sin(\omega t)$ applied in the area occupied by the electron. The Fig.~\ref{fig5} shows the motion of the wavepacket subjected to such a field. In the time interval between $t=0$~ps and $t=300$~ps the expectation value of the position (blue curve) oscillates with a linearly increasing amplitude. The energy (red curve) rises quadratically in a step-wise manner, because the wavenumber rises by the same value with each oscillation of position. These results are in analogous to the classical harmonic oscillator with resonant driving where amplitude  linearly  increases  over  time.

An increase in the position oscillations amplitude, and hence also in energy, requires an exactly parabolical confinement potential. Non-parabolicity constitutes a natural limit to energy growth. In the simulation shown in Fig.~\ref{fig5}, the wavepacket is confined in a parabolic potential but the calculations are performed on a finite fragment of the quantum wire which effectively imposes infinite potential walls at both ends of the wire. As the electron approaches the wire borders, the frequency of its oscillations is no longer consistent with the frequency of the time-varying electric field inducing its movement. The amplitude of spatial oscillations of the wavepacket ceases to grow. The influence of nonparabolicity is easily visible in Fig.~\ref{fig5} for $t>300$~ps.

In all the presented simulations we assumed that initially the electron spin is parallel to the $z$-axis. Despite the electron movement, the spin did not change, since according to the Hamiltonian (Eq. \ref{hso}), movement along the $x$-axis implies rotation of spin around the $z$-axis. The situation would be similar if the initial spin was antiparallel to the $z$-axis. It would only result in an opposite direction for the electron movement. 

\subsection{Spin density separation}
Now let us assume that the spin of the electron is neither parallel nor antiparallel to the $z$-axis, but is a linear combination of both basis vectors (Eqs. \ref{groundstate1}, \ref{groundstate2}), forming the two-row spinor:
\begin{equation}\label{spinor}
\Psi(x,t)=\begin{pmatrix}\psi_{\uparrow}(x,t)\\\psi_{\downarrow}(x,t)\end{pmatrix}.
\end{equation}
As mentioned before, after the SOI is turned on, the upper and lower parts of the spinor gain opposite momenta; hence, this effect can be used for spatial spin separation.

This time, we initially set up the electron spin as an equally weighted linear combination of spins up and down. Moreover, the electron is trapped in a parabolic potential and we initialize it as the ground state of the harmonic oscillator with an assumption of no SOI ($E_y=0$); thus, the wavefunction is of the form:
\begin{equation}
\Psi(x,t=0)=\left(\frac{\beta}{2\pi}\right)^\frac{1}{4}\begin{pmatrix}1\\1\end{pmatrix}e^{-\beta x^2}.
\end{equation}

\begin{figure}[b]
\centering
\includegraphics[width=0.45\textwidth]{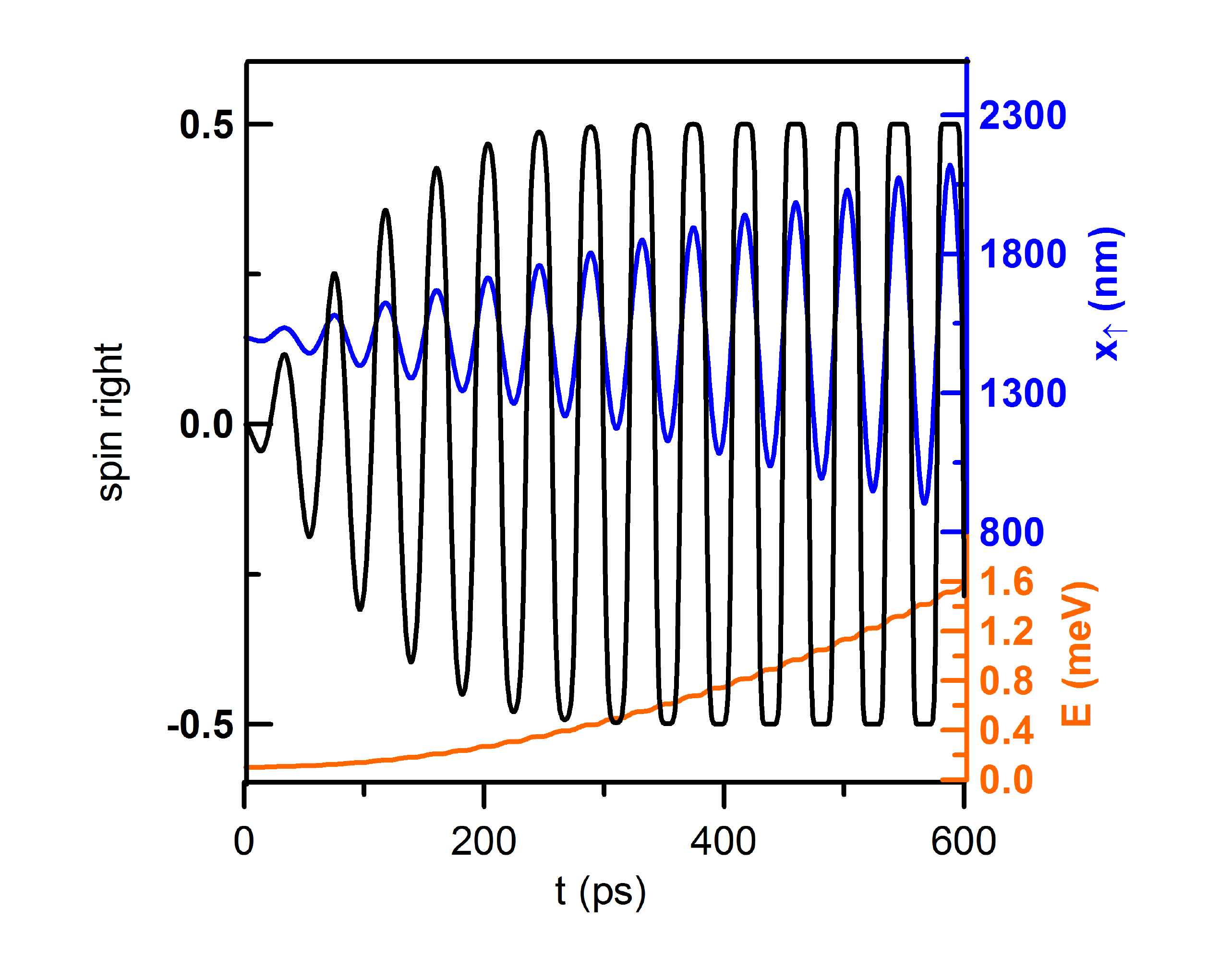}
\caption{\label{fig6}Time evolution of the spin density of the electron trapped in a harmonic potential with sinusoidally varying SOI. Initially the spin is set as an equally weighted linear combination of spin up and down. The blue curve denotes the expectation value of the position $x_{\uparrow}(t)$ calculated for the upper half of the spinor, the red curve---the electron energy, and the black curve---the expectation value of spin $\sigma_z^\mathrm{R}(t)$ in the right half of the quantum wire.}
\end{figure}
Next, we apply an electric field varying sinusoidally with the frequency tuned to the eigenfrequency of the chosen harmonic potential. Both parts of the spinor behave in different ways, such as wavefunctions of the electron with spin up and down (Eq. \ref{groundstate1}, \ref{groundstate2}). The center of mass of the entire electron does not move; however, if calculated for each spinor component respectively, they move away in an oscillatory fashion with opposite phases and growing amplitude. In Fig.~\ref{fig6}, we see the time evolution of such a system. The expectation value of position for the spin up (blue curve) is calculated using the upper component of the spinor (Eq. \ref{spinor}) as
\begin{equation}
x_\uparrow(t)=\frac{\int_0^Lx|\psi_{\uparrow}(x,t)|^2\,\mathrm{d}x}{\int_0^L|\psi_{\uparrow}(x,t)|^2\,\mathrm{d}x},
\end{equation}
with $L$ being the length of the quantum wire.
The oscillations of the center of mass of the electron density with spin down $x_\downarrow(t)$ (not presented) are similar, but with a phase shifted by $\pi$. The expectation value of the Pauli-$z$ matrix $\hat{\sigma}_z$ (black curve), calculated in the right half of the quantum wire, is
\begin{equation}
\sigma_z^\mathrm{R}(t)=\left\langle\Psi\right|\hat{\sigma}_z\left|\Psi\right\rangle_\mathrm{R}=\int_{L/2}^{L}\left(|\psi_{\uparrow}(x,t)|^2-|\psi_{\downarrow}(x,t)|^2\right)\!\mathrm{d}x
\end{equation}
and denoted by \textit{spin right} in the figures. 

The expectation position of the upward spin density $x_{\uparrow}(t)$ oscillates similar to the expectation value of position of the electron from Fig.~\ref{fig5}. In the parabolic region of the potential, the amplitude of oscillations grows linearly with time. As the oscillations of spin up and down wavefunctions are phase-shifted by exactly $\pi$, these components separate. Now let us look at the orientation of spin in the right half of the quantum wire $\sigma_z^\mathrm{R}(t)$ (black curve). The amplitude stops growing as it reaches the value $0.5$ (or $-0.5$), meaning that the entire spin in the right half is oriented upwards (downwards). 
For better visualization we can define the spatial $z$-spin component density as
\begin{align}
&\rho_{\sigma}(x,t)=\Psi^{\dag}(x,t)\hat{\sigma}_z\Psi(x,t)= \nonumber\\
&=\big(\psi^\ast_\uparrow(x,t),\psi^\ast_\downarrow(x,t)\big)\!\begin{pmatrix}1 & 0\\0 & -1\end{pmatrix}\! \begin{pmatrix}\psi_\uparrow(x,t) \\ \psi_\downarrow(x,t) \end{pmatrix}= \nonumber\\
&=|\psi_{\uparrow}(x,t)|^2-|\psi_{\downarrow}(x,t)|^2,
\end{align}
and compare it with the total electron density defined as
\begin{equation}\label{chargedist}
\rho(x,t)=\Psi^{\dag}(x,t)\Psi(x,t)=|\psi_{\uparrow}(x,t)|^2+|\psi_{\downarrow}(x,t)|^2.
\end{equation}

\begin{figure}[t]
\centering
\includegraphics[width=0.45\textwidth]{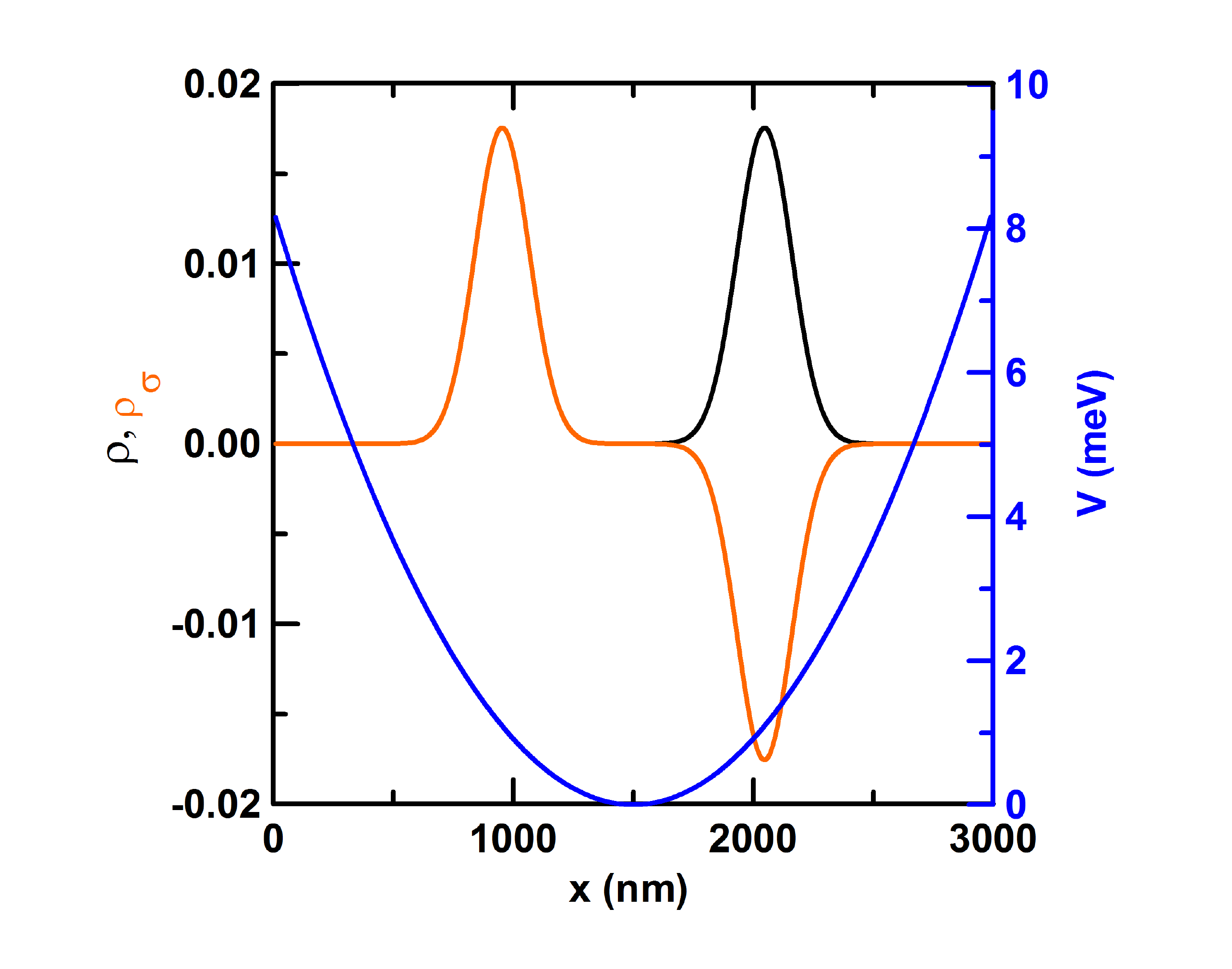}
\caption{\label{fig7}Comparison of the spatial $z$-spin density $\rho_{\sigma}(x,t)$ (red curve) with the total spatial density $\rho(x,t)$ (black curve) at the moment of maximal separation. The spin density divides into two separate parts with opposite spins. Additionally, the confinement potential is shown as a blue curve. }
\end{figure}

Fig.~\ref{fig7} shows a comparison of the spin density (red curve) and the total density (black curve) at the moment of maximal spin separation (spin in the right half reaching $-0.5$).
In the left part of the wire the spin density completely covers the total density, meaning that in this region the spin is oriented upwards. Consequently, the right half contains the spin oriented downwards. 
Now, if we put a potential barrier in the center of the quantum wire, the spin density divides into two spatially separated parts with opposite spins. 

\begin{figure}[b]
\centering
\includegraphics[width=0.45\textwidth]{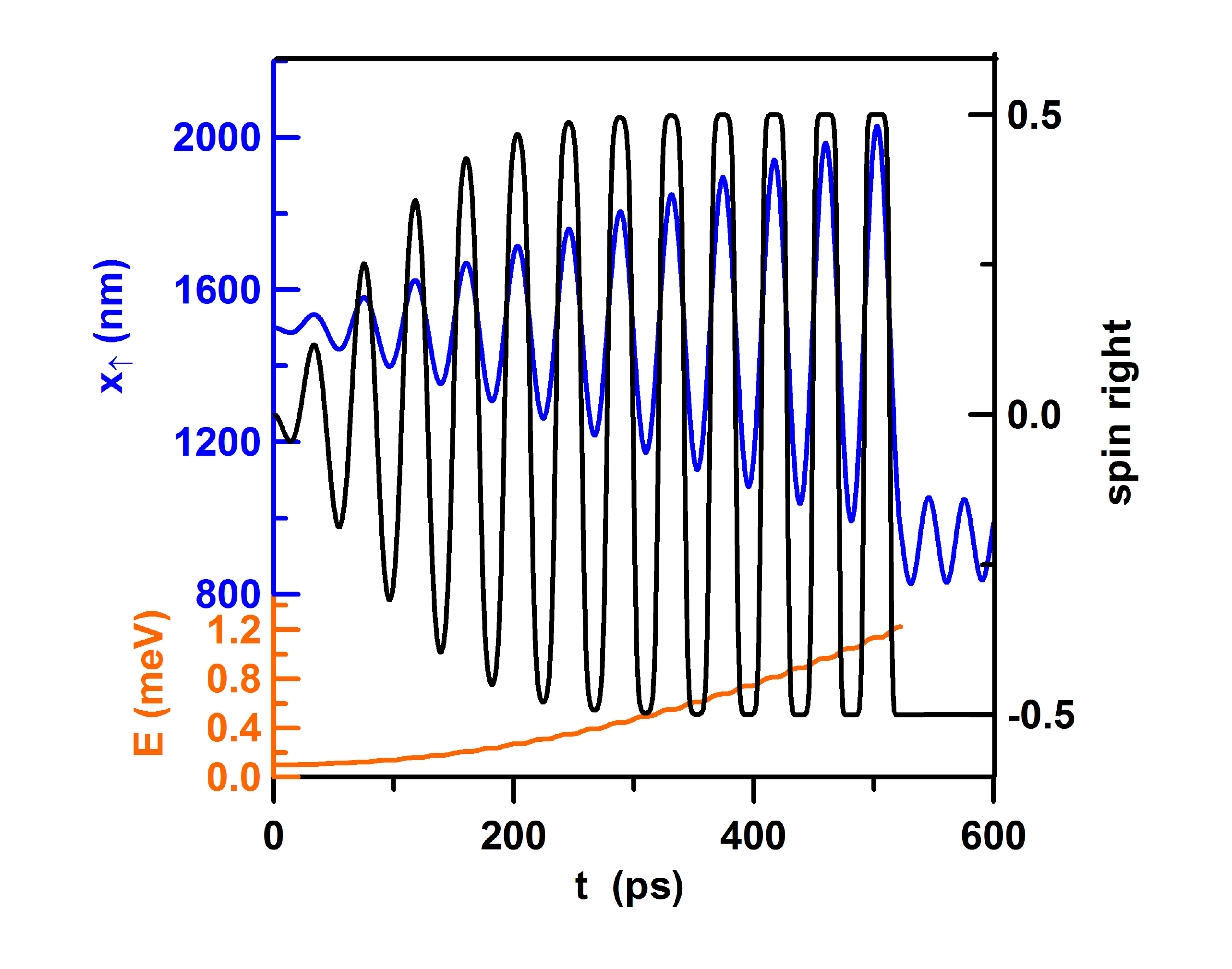}
\caption{\label{fig8}Simulation of the time evolution of the electron spin (as in Fig.~\ref{fig6}), but with the creation of a potential barrier at about $t=500$~ps with simultaneous cessation of the RSOI oscillations. Markings as in the Fig.~\ref{fig6}.}
\end{figure}

We should also notice that, as the amplitude of oscillation of the spin up position $x_\uparrow(t)$ increases, the course of the  spin in right half of the wire $\sigma_z^\mathrm{R}(t)$ (black curve in the Fig.~\ref{fig6}) starts to form plateaus and resemble a square wave. Therefore, time intervals with separated spins (plateaus) become longer compared to the intervals of spin changes. This increases the tolerance for the moment of the potential barrier creation.  
Fig.~\ref{fig8} shows the results of a simulation, in which a potential barrier was created in the center of the potential well at about $t=500$~ps.  

The wavefunction of the electron state shown in Fig.~\ref{fig7} is of Schr\"odinger's cat type \cite{winelandzrodlo, winelandprzeglad} and it can be written as $\left|\uparrow\right\rangle\left|L\right\rangle+\left|\downarrow\right\rangle\left|R\right\rangle$ with states $\left|\uparrow\right\rangle$ and $\left|\downarrow\right\rangle$ denoting spin orientations, and states $\left|L\right\rangle$ and $\left|R\right\rangle$ denoting localization of the electron, respectively, in the left and right half of the quantum wire. 
After separation of both parts of the spin density with a potential barrier, the parts can be relocated to arbitrary positions and the wavefunction still describes the Schr\"odinger's cat state. Such a wavefunction also has another important feature, that is, it constitutes an entangled state of spin state and state localized in two different spatial regions \cite{winelandprzeglad}. This state has been observed experimentally in ion traps \cite{winelandzrodlo, wineland1, wineland2, inni}.
The purpose of our study is to design a nanodevice based on a planar semiconductor heterostructure in which this effect could be observed. 

In simulations presented in Figs.~\ref{fig1}-\ref{fig5}, we have shown movement of the electron as a whole. In this case, self-focusing was beneficial, since it stabilizes the shape of the wavepacket. Its influence was calculated with the image charge method under the assumption that distance $d$ between the quantum wire and the metallic gate was equal to $50$~nm. 
In the subsequent simulations (Figs.~\ref{fig6}-\ref{fig8}) the self-focusing effect disturbs the spin separation process and had to be eliminated. During separation both parts of opposite spins interact with each other through the induced charge. The interaction potential is nonparabolic and destroys the parabolicity of the total confinement potential effectively changing the frequency of oscillations. 
The self-focusing was mitigated by placing the quantum wire at a greater distance from the gate (we assumed $d=1000$~nm). During the design stage of a real nanodevice we cannot proceed this way and self-focusing has to be compensated in a different manner.

\section{Nanodevice generating Schr\"odinger's cat states\label{secondpart}}

\subsection{Nanodevice design and principles of simulation}
\begin{figure}[b]
\hspace{-7mm}
\includegraphics[width=0.5\textwidth]{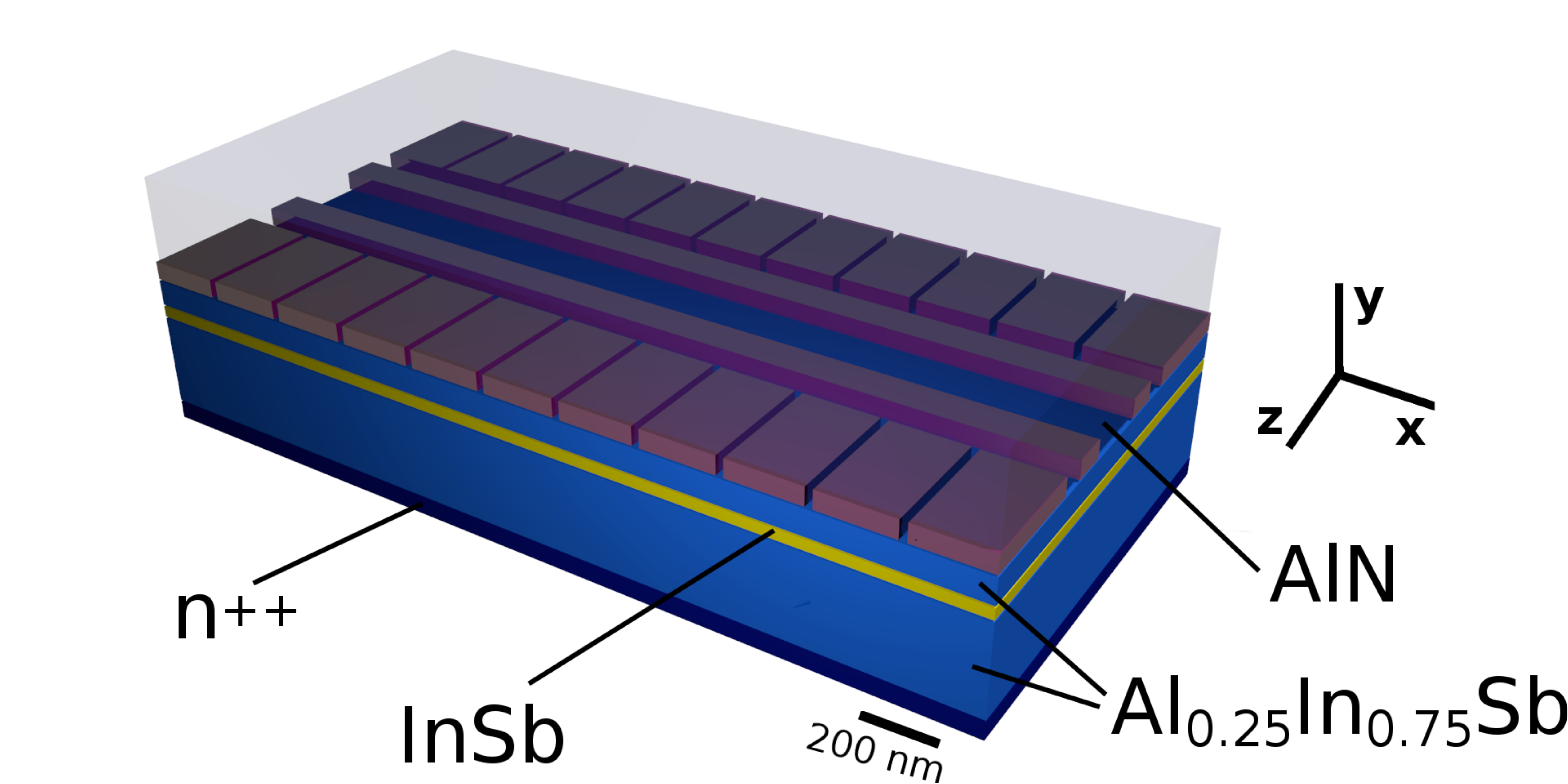}
\caption{\label{fig9}Schematic view of the proposed multi-gated nanodevice for the spin density separation, which leads to the generation of the Schr\"odinger's cat states. The top gate is not shown.}
\end{figure}

\begin{figure}
\hspace{-1.2cm}
\includegraphics[width=0.4\textwidth]{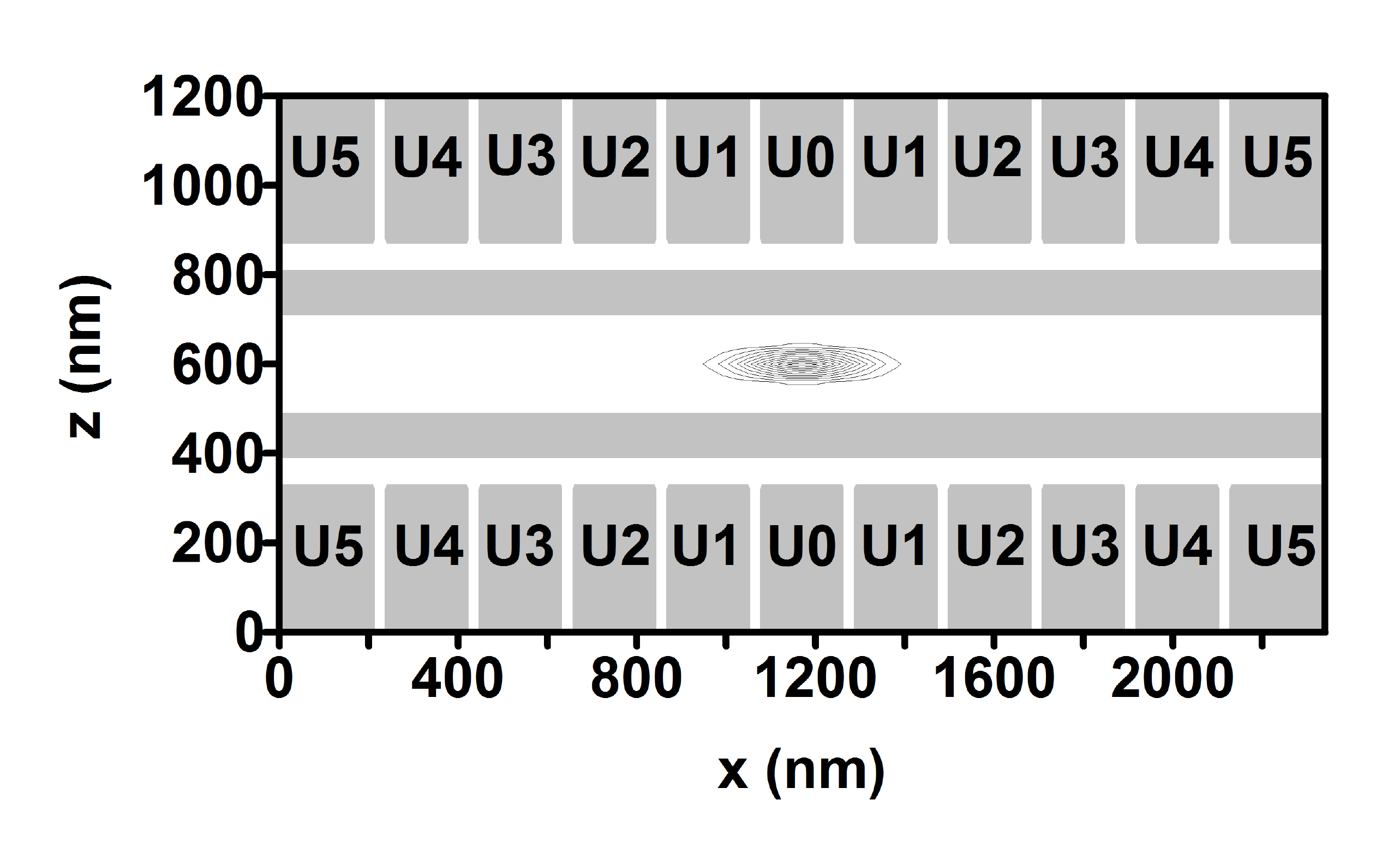}
\caption{\label{fig10}The layout of the gate array viewed from above: the long rails between pairs of the side electrodes; the top gate (covering whole device) is not visible. The density of the trapped electron is also shown, localized at the initial moment in the center, calculated as $|\Psi(x,z,t=0)|^2$.}
\end{figure}

For practical realization of the nanodevice we propose a typical gated planar semiconductor heterostructure grown in the $y$ direction, with a quantum well (QW) parallel to its surface ($x$,$z$ plane). The Fig.~\ref{fig9} shows a schematic view of the proposed nanodevice.
The QW is made of $\mathrm{InSb}$ which has a relatively high Rashba coupling. Two blocking layers, below and above the QW, are made of $\mathrm{Al_xIn_{1-x}Sb}$ ternary with $x=25$~$\%$, for which the bottom of the conduction band is shifted up by $320$~meV, creating potential barriers \cite{offset1, offset2, offset3}. The substrate may consist of $\mathrm{Al_xIn_{1-x}Sb}$ highly doped with donors (n$^{++}$).
The lower blocking layer is $230$~nm thick, while the upper one is of a thickness of $50$~nm. The InSb QW inbetween is $20$~nm thick. An array of gates, as depicted in Fig.~\ref{fig10} ($y_0$ denotes the $y$-position of the QW), is deposited on the heterostructure.
Similar, somewhat complex, gate layouts can be found in various experimental works\cite{st1,coupled,st3,st4}, 
being successfully deposited on a surface of a heterostructure.
Moreover, the top of the structure is covered with a layer of dielectric material (AlN) of thickness $320$~nm. Finally, we put an additional layer of metal on the dielectric (not shown in the picture).

We choose InSb for the quantum well material, as it is characterized by a large value of spin-orbit coupling $\alpha_\mathrm{so}$, which allows for effective generation of cat states in times much shorter than the coherence time of the spin-qubit. The choice of InAs with a several times smaller value of $\alpha_\mathrm{so}$ would increase several times the time of operation of the device.

If there is no electric field, the electrons fill the QW up to the Fermi level, creating a two dimensional electron gas (2DEG). Electrons trapped in the QW have two degrees of freedom in lateral directions ($x$ and $z$). By applying negative voltages to gates (in relation to the substrate) we can remove the 2DEG from the quantum well leaving only one electron trapped between two long central gates (rails). The voltages applied to the rails block electron movement along the $z$-axis. To the remaining gates, we apply voltages creating a parabolic confinement potential along the $x$-axis.

The potential distribution in the nanodevice is calculated by solving the Poisson's equation in a computational box surrounding the entire device. To simplify the boundary conditions setup, we covered the entire structure with a layer of metal, although it does not have any considerable impact on the operation of the nanodevice. We have chosen a computational box with dimensions $L_x=2340$~nm, $L_z=1200$~nm, $L_y=620$~nm as optimal. 
In the simulations we consider all the complexity of the structure by including the geometry details, voltages applied to the substrate and gates, the time-dependent distribution of the electron in the QW, and also the charge induced on the gates or the (conducting) substrate. 

The confined electron is treated as a particle in a 2D QW with a motion frozen in the $y$ direction; hence, we assume a time-dependent Hamiltonian of the form:
\begin{equation}
\begin{split}\label{realham}
H(x,z,t)&=\left(\!-\frac{\hbar^2}{2m}\left(\partial^2_x+\partial^2_z\right)-|e|\varphi_{\mathrm{elst}}(x,z,y_0,t)\!\right)\!1_2\\
&+H_\mathrm{R}(x,z,y_0,t).
\end{split}
\end{equation}
Here $m=0.014~m_e$ denotes the effective band mass of an electron in $\mathrm{InSb}$ material, and $\varphi_{\mathrm{elst}}$ constitutes the potential felt by the electron calculated as
\begin{equation}
\varphi_{\mathrm{elst}}(\mathbf{r},t)=\varphi_{\mathrm{tot}}(\mathbf{r},t)-\varphi_{\mathrm{self}}(\mathbf{r},t),
\end{equation}
with $\varphi_{\mathrm{tot}}$ being the total potential inside the nanodevice calculated at each time step using the generalized Poisson's equation
\begin{equation}
\nabla\cdot\left[\epsilon_0\epsilon(\mathbf{r})\nabla\varphi_{\mathrm{tot}}(\mathbf{r},t)\right]=-\rho(\mathbf{r},t).
\label{pois}
\end{equation}
This is solved with boundary conditions created by voltages applied to the gates. The charge distribution $\rho(\mathbf{r},t)$ is calculated as in (Eq. \ref{chargedist}), but here in three dimensions. We also need to subtract the Coulomb potential, originating directly from the electron distribution, from the total potential to avoid electron self-interaction.
This is calculated as
\begin{equation}
\varphi_{\mathrm{self}}(\mathbf{r},t)=\frac{-|e|}{4\pi\epsilon\epsilon_0}\int\mathrm{d}^3\mathbf{r}^\prime\frac{\rho(\mathbf{r,t}^\prime)}{|\mathbf{r}-\mathbf{r}^\prime|}.
\end{equation}
Details of this method can be found in [\onlinecite{metoda}]. The last term of (Eq. \ref{realham}) constitutes the RSOI given by the following Hamiltonian\cite{bychkov,rashba2,winkler}:
\begin{equation}
H_\mathrm{R}(\mathbf{r},t)=-\frac{\alpha_\mathrm{\mathrm{so}}|e|}{\hbar}\left[\nabla\varphi_{\mathrm{elst}}(\mathbf{r},t)\times\mathbf{p}^\mathrm{\scriptscriptstyle 2D}\right]\cdot\boldsymbol{\sigma},
\label{raszka}
\end{equation}
with two-dimensional electron momentum $\mathbf{p}^\mathrm{\scriptscriptstyle 2D}=\left(p_x, 0, p_z\right)$ and the Rashba coupling $\alpha_{\mathrm{\mathrm{so}}}=5.23$~$\mathrm{nm}^2$ for $\mathrm{InSb}$ material\cite{winkler}. The Pauli vector $\boldsymbol{\sigma}=(\sigma_x,\sigma_y,\sigma_z)$.

The value of the Rashba coupling strictly depends on quantum well material (InSb) parameters.
It may also depend on the barrier material and details of the electron confinement within the heterostructure in $y$ direction. The exact coupling calculations and the validity of the used Rashba model have been included in the appendix.
If the InSb layer is grown in the [111] crystallographic direction, 
the Dresselhaus spin-orbit interaction Hamiltonian\cite{dr1,hanson,dr3} 
can be reduced to a pure Rashba-like term: $\sim\!p_x\sigma_z$.\cite{dr44} 
Such a term adds a constant offset to the Rashba coupling and does not affect 
the presented spin density separation process.

We apply voltages to the gate array depicted in the Fig.~\ref{fig10}. Initially the rails and the top gate (not visible) are set to $U=-400$~mV. To the side gates located in the center of the structure (gates $\mathrm{U_0}$) we apply voltages $10$~mV lower ($U_0=U-10\,\mathrm{mV}=-410$~mV). Voltages for the remaining side gates are calculated as $U_i=U_0-\gamma i^2$ with $\gamma=5$~mV. The zero reference voltage $U_\mathrm{ref}=0$ is applied to the doped substrate.

Now, we generate the ground state of an electron using the imaginary time evolution method \cite{czasur}. 
Then, during the real time evolution of the system we change sinusoidally the voltages applied to all the gates, according to the formula 
\begin{equation}\label{voltvar}
U(t)=\tilde{U}+\Delta \tilde{U}\sin(\omega t),
\end{equation}
with $\tilde{U}=-400$~mV, $\Delta \tilde{U}=300$~mV and $\omega$ tuned to the characteristic frequency of the obtained harmonic confinement potential. 
Specifically, $U(t)$ depicts the voltage applied to the rails and the top gate.
Voltages applied to the remaining (side) gates are shifted in relation to the $U(t)$ voltage in the 
same way as at the beginning of the simulation. 
Consequently, the shape of the confinement potential along the $x$-axis remains virtually the same 
(parabolic with the same curvature) for the entire simulation, 
even though the potential is shifted by a time-dependent value. 
Oscillations of the gate voltages (Eq.~\ref{voltvar}), relative to zero voltage on the substrate, cause oscillations of 
the electric field ($\mathbf{E}_y$) perpendicular to the surface of the heterostructure, and thereby the SOI oscillates as well.
Thus, we get effectively  $H_\mathrm{R}(t) \sim \alpha_\mathrm{so} E_y(t)\,p_x\sigma_z$ (for $x$ motion).

The electric field modulating the spin-orbit interaction is parallel to the growth direction ($y$-axis). This combined with electron motion along the $x$-axis gives a spin quantization axis as $z$. Our choice of the growth direction is motivated by a widely used convention to make the $z$ direction as the spin quantization axis.

\subsection{Preliminary simulations}
\begin{figure}
\centering
\includegraphics[width=0.45\textwidth]{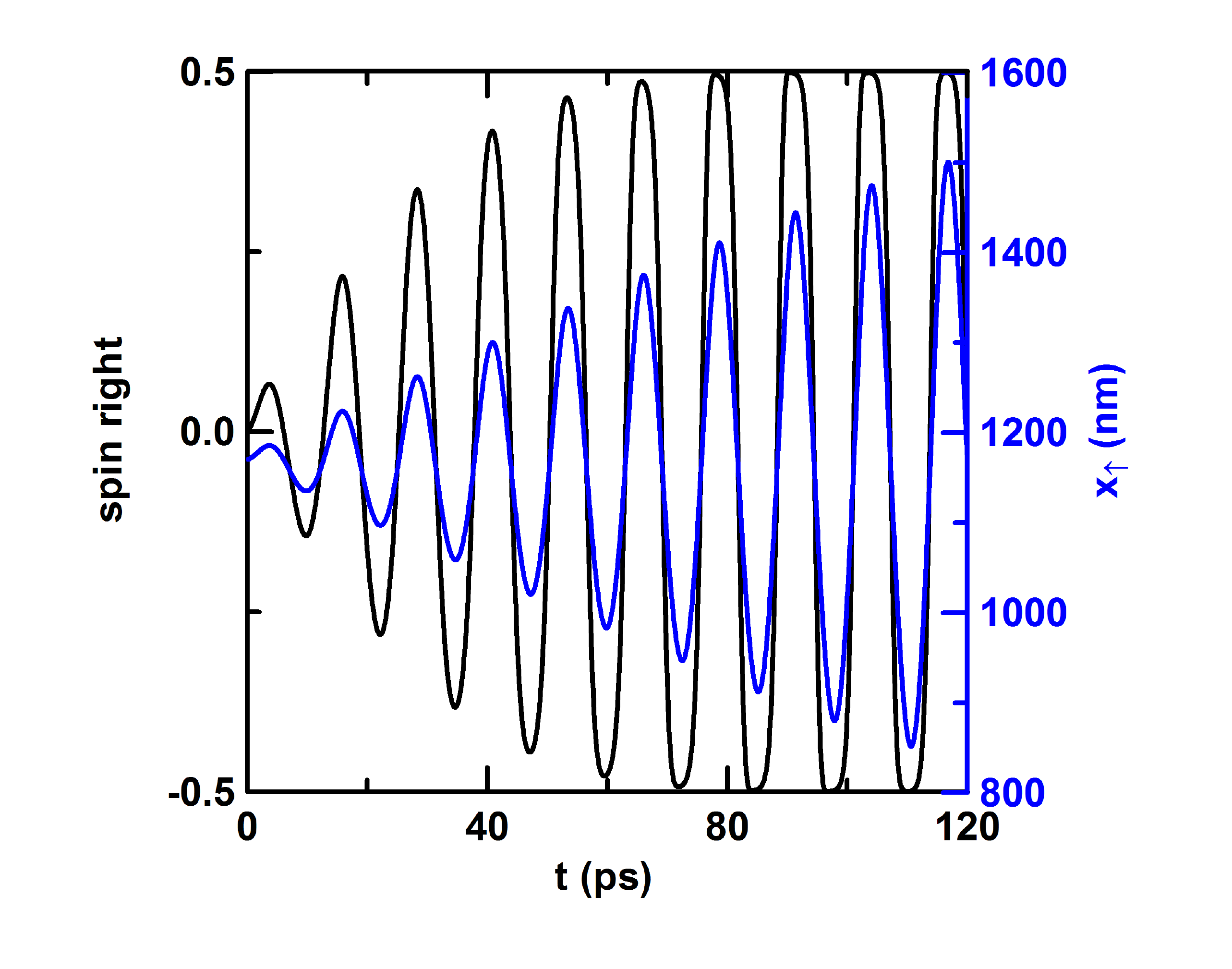}
\caption{\label{fig11}Simulation of the time evolution for the electron trapped in the nanodevice presented in the Figs.~\ref{fig9} and \ref{fig10} with sinusoidally varying voltages applied to the gates. 
Initially the electron spin is oriented along the $x$-axis.
The blue curve shows the position calculated for the upper spinor part $x_\uparrow(t)$ and the black one denotes the $z$-spin component calculated in the right half of the potential well $\sigma_z^{\mathrm{R}}(t)$. This simulation neglects the self-focusing effect.}
\end{figure}

Since the simulation within the 1D model has shown that self-focusing destroys the potential parabolicity, the first 3D simulations were performed without this effect. This can be easily done by neglecting the presence of charge in the Poisson's equation (\ref{pois}). In Fig.~\ref{fig11}, we see results of such a simulation obtained for a frequency of oscillations (Eq. \ref{voltvar}) corresponding to the harmonic oscillator energy $\hbar\omega=0.331\,\mathrm{meV}$.

We start from the electron wavefunction with spin oriented along the $x$-axis, i.e. an equally weighted linear combination of spin up and down components, generated as the ground state of the parabolic confinement potential. We perform simulations with a correctly tuned voltage oscillation frequency (in Eq. \ref{voltvar}). As a result, we get expectation values of spin in the right half of the nanodevice (black curve) similar to the time courses from Fig.~\ref{fig6}. The curve reaches the value $\sigma_z^\mathrm{R}(t)>0.499$, which indicates almost full spatial spin separation. The long plateau regions with $\sigma_z^\mathrm{R}(t)\approx 0.5$ mean that the intervals in which the spin remains separated are long enough to set an additional potential barrier between the separated spin density parts to further enhance separation. 

The period of oscillations of voltages applied to the gates of the order of $10$~ps is near the limits of current technological capabilities. However, we should note that the frequency of oscillations depends on the parabolic confinement potential and can be reduced several times by lowering the control voltages $U_i$ by reducing the $\gamma$ coefficient.

\begin{figure}[b]
\centering
\includegraphics[width=0.45\textwidth]{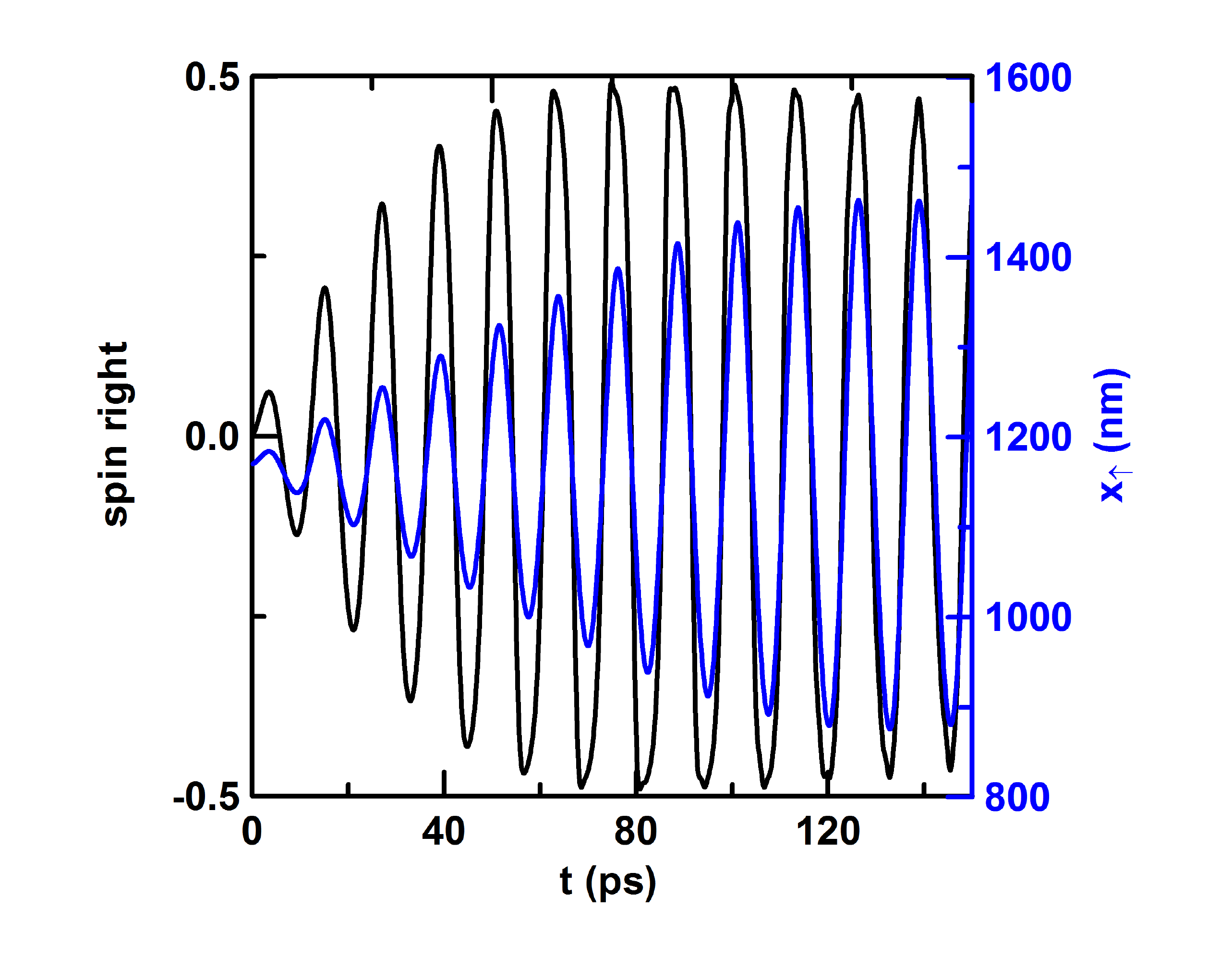}
\caption{\label{fig12}Simulation similar to that from Fig.~\ref{fig11} but with  self-focusing effect included and the permittivity value for $\mathrm{InSb}$ ($\epsilon=17.9$) set for the entire computational box. The markings are as in Fig.~\ref{fig11}.}
\end{figure}
In Fig.~\ref{fig12}, we can see another case of such simulations, but this time with the self-focusing effect included. 
The Poisson's equation is solved with the same permittivity ($\mathrm{InSb}$) $\epsilon=17.9$ for the entire computational box. This corresponds to a hypothetical situation in which the nanostructure is covered not with a dielectric layer but with a semiconductor layer of permittivity similar to the $\mathrm{InSb}$. This time the wavefunctions with opposite spins do not fully separate and $\sigma_z^{\mathrm{R}}(t)$ does not reach the value of $0.5$ (or $-0.5$) closely enough. The Poisson's equation solution includes the self-focusing effect caused by the induced charge in both, the gates and the conducting substrate. The sources of this charge have an influence on the potential felt by the electron.  

\begin{figure}[b]
\centering
\includegraphics[width=0.45\textwidth]{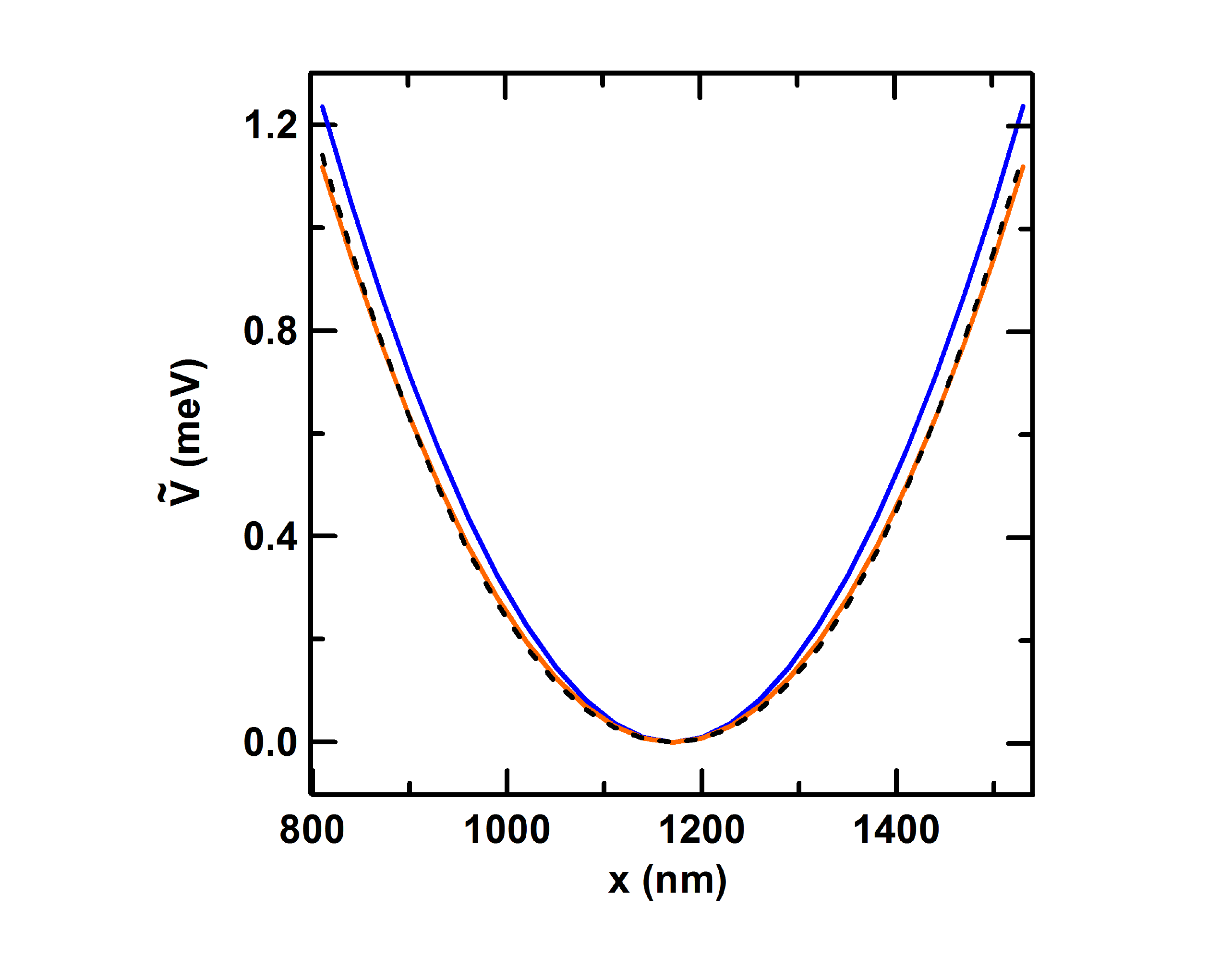}
\caption{\label{fig13}Potential energies in the vicinity of their minima, calculated for three cases: (red curve) with self-focusing neglected, (blue curve) with self-focusing taken into account, and (black dashed curve) with self-focusing included for a nanostructure covered with a dielectric layer. All potentials are shifted, so that the minima coincide.}
\end{figure}

Let us look at Fig.~\ref{fig13}.
It presents how the potential energy in the QW changes with the inclusion of the self-focusing effect. The case with self-focusing (blue curve) clearly deviates from the case without self-focusing (red curve). The potential energy for the first case (blue curve) remains acceptably parabolical, but during wavefunction separation, the interaction with the induced charge weakens, since the wavepacket splits into two separated parts and the potential felt by each one changes. These changes and the resulting non-parabolicity hinder the electron acceleration. The effect of self-focusing could be mitigated by increasing the distance between the QW and the gates or the substrate. 

\subsection{Effect of adding a dielectric}
In the proposed nanostructure (Fig.~\ref{fig9}), the distance between the QW and the substrate is $240$~nm and for the top metallic gate $380$~nm, which are high enough for negligible self-focusing. However, to neutralize the influence of self-focusing originating from the rails and the side gates, we employ another nanodevice design. This is achieved by placing a dielectric layer on top of these gates---see Fig.~\ref{fig9} (but under the top gate), with permittivity lower than the permittivity of the QW material. For this purpose, we used $\mathrm{AlN}$ material with permittivity $\epsilon_\mathrm{AlN}=7.6$ ($\epsilon_\mathrm{InSb}=17.9$).
If we simply used the image charge method for the interface of two dielectric media, the image charge induced within the material of lower permittivity $-q(\epsilon_\mathrm{AlN}-\epsilon_\mathrm{InSb})/(\epsilon_\mathrm{AlN}+\epsilon_\mathrm{InSb})\simeq0.4\,q$ would have the same (negative) sign as the primary charge $q=-|e|$ (located in the material of higher permittivity). Therefore, such a dielectric addition could compensate the (positive) charge induced on the metallic gates.

We observe such an effect in simulations. In Fig.~\ref{fig13}, the black dashed curve shows the potential energy calculated for one such case of compensation. 
In the device region between $800$~nm and $1500$~nm in which the electron is accelerated and separated ($x_\uparrow(t)$ oscillations range---see: Fig.~\ref{fig12}), the black curve from Fig.~\ref{fig13} coincides with the red one.
This clearly means that self-focusing has been neutralized by defocussing due to the presence of the dielectric layer. Such a compensation also requires a careful tuning of the rail widths and the distance between them. Fig.~\ref{fig10} shows relatively large areas not covered with any gates. In these areas the semiconductor comes into direct contact with the dielectric. 

\begin{figure}[t]
\centering
\includegraphics[width=0.45\textwidth]{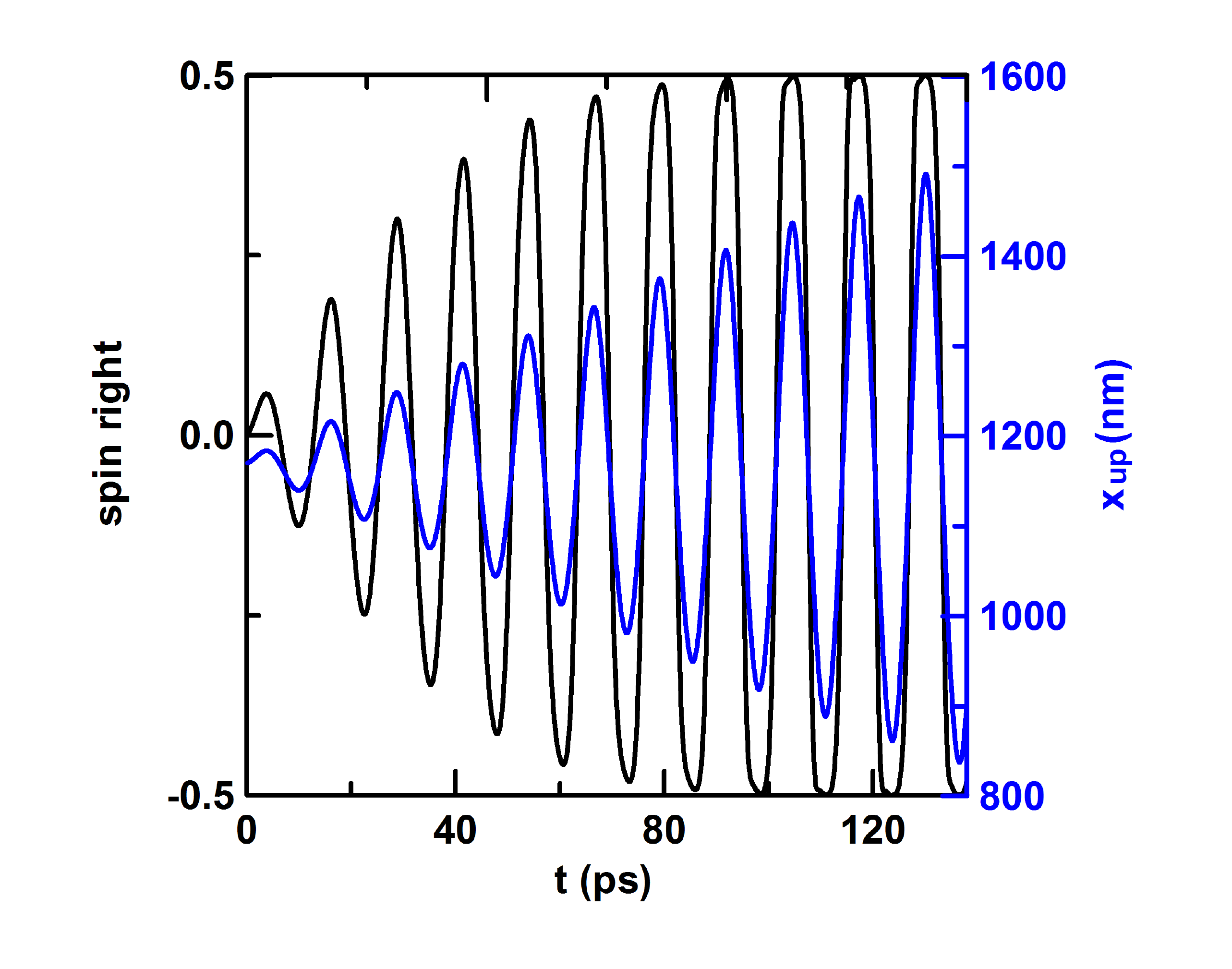}
\caption{\label{fig14}Simulation of the time evolution of the electron trapped in the nanostructure with dielectric compensation of self-focusing. Markings as in Fig.~\ref{fig11}.}
\end{figure}
Fig.~\ref{fig14} shows results of the simulation with spin-dependent electron acceleration due to oscillating spin-orbit coupling in the nanostructure covered with the dielectric layer. This time the tuned voltage oscillation frequency is of $\hbar\omega=0.325\,\mathrm{meV}$.   
The course of the expectation value of the spin right $\sigma_z^{\mathrm{R}}(t)$ starts to resemble ones from Figs.~\ref{fig6} and \ref{fig11}. Characteristic plateaus appear again, giving a lot of time for setting up a potential barrier between the separated parts. The most favorable moment to create such a barrier is at $t_1=123$~ps. At this moment we stop any oscillations of voltages and keep them fixed. Moreover, for the two gates marked as $\mathrm{U_0}$ the voltages are decreased by $150$~mV, namely $U_0(t_1)=U(t_1)-160$~mV. This procedure elevates the potential in the center of the structure, creating a potential barrier. This effectively divides the nanodevice into two regions: left and right. 
\begin{figure}[b]
\centering
\includegraphics[width=0.45\textwidth]{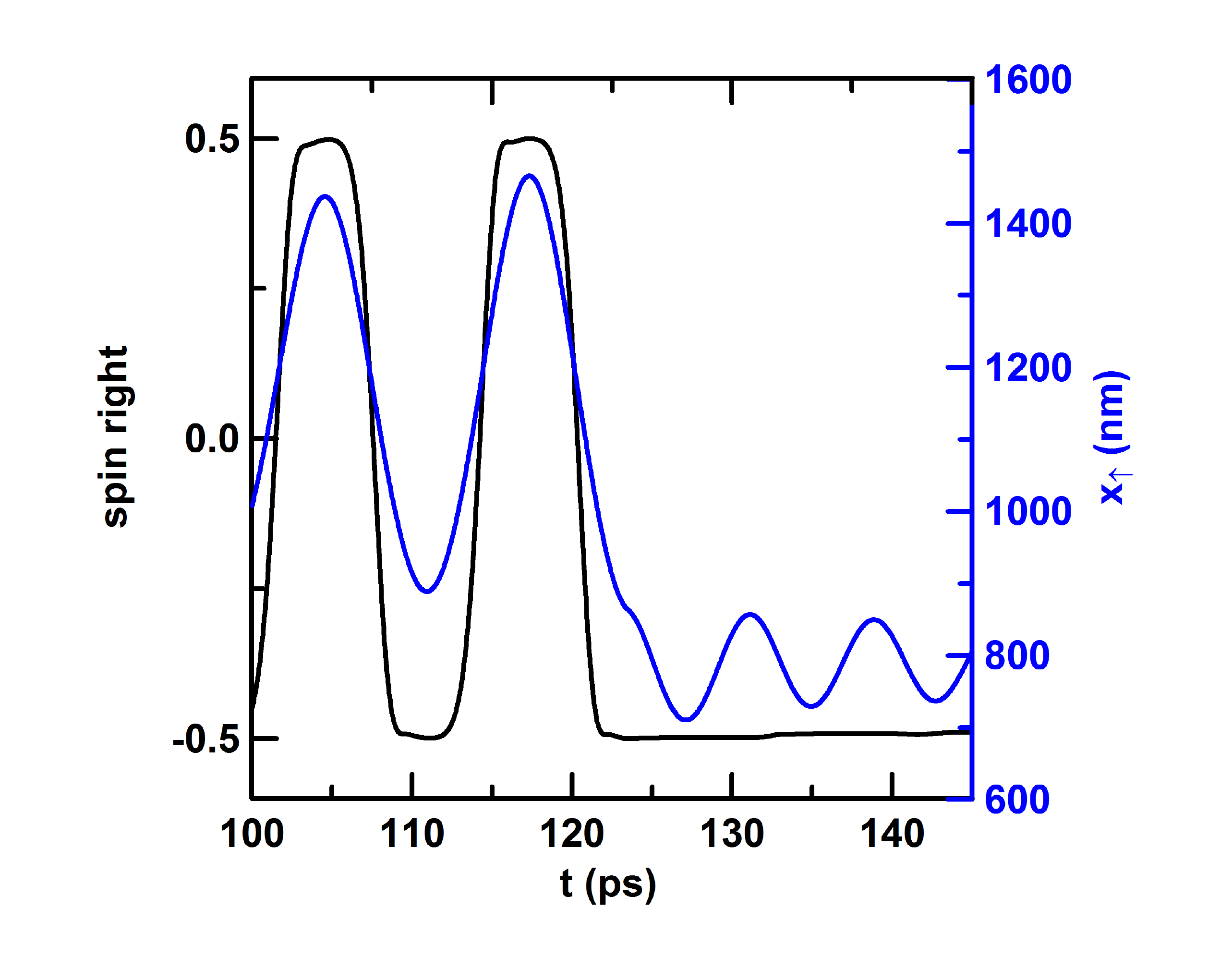}
\caption{\label{fig15}Fragment of the  time evolution with visible results of setting up the barrier, dividing the nanodevice into two regions. Markings as in Fig.~\ref{fig11}.}
\end{figure}
The last fragment of the simulation depicting the barrier setup moment is shown in Fig.~\ref{fig15}. The $\sigma_z^{\mathrm{R}}(t)$ (black curve) remains close to the value $-0.5$, which means that spin in the right part of the nanodevice is oriented down. The spin up position $x_\uparrow(t)$ (blue curve) no longer returns to the upper part of the graph but it falls and remains down due to the repulsive influence of the barrier. 

Now, the $x_\uparrow(t)$ starts to oscillate in the left half of the nanodevice. These oscillations indicate that, during the separation stage, the electron did not get rid of the excess of energy. The frequency of the oscillations has also changed, since the local curvature of the confinement potential is now slightly different. The oscillations have a negative influence on the nanodevice operation, because the separated spin density part
can tunnel through the barrier while colliding with it. This manifests itself via a small rise in the final part of the $\sigma_z^{\mathrm{R}}(t)$ course (barely visible in Fig.~\ref{fig15}). 

\begin{figure}[t]
\centering
\includegraphics[width=0.45\textwidth]{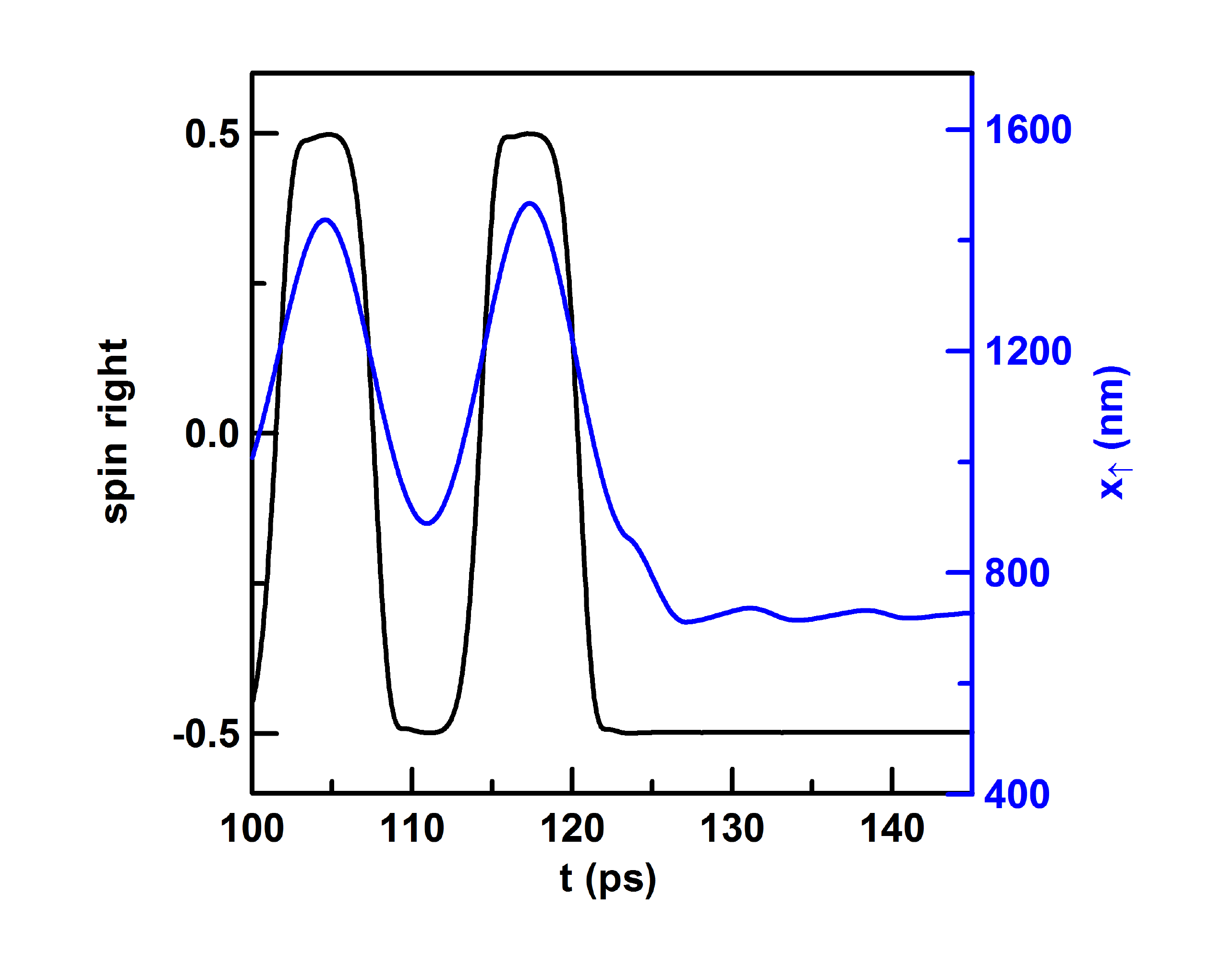}
\caption{\label{fig16}The final fragment of the time evolution of the system with the potential barrier setup and its subsequent elevation. Markings as in Fig.~\ref{fig11}.}
\end{figure}

\begin{figure}[t]
\centering
\includegraphics[width=0.45\textwidth]{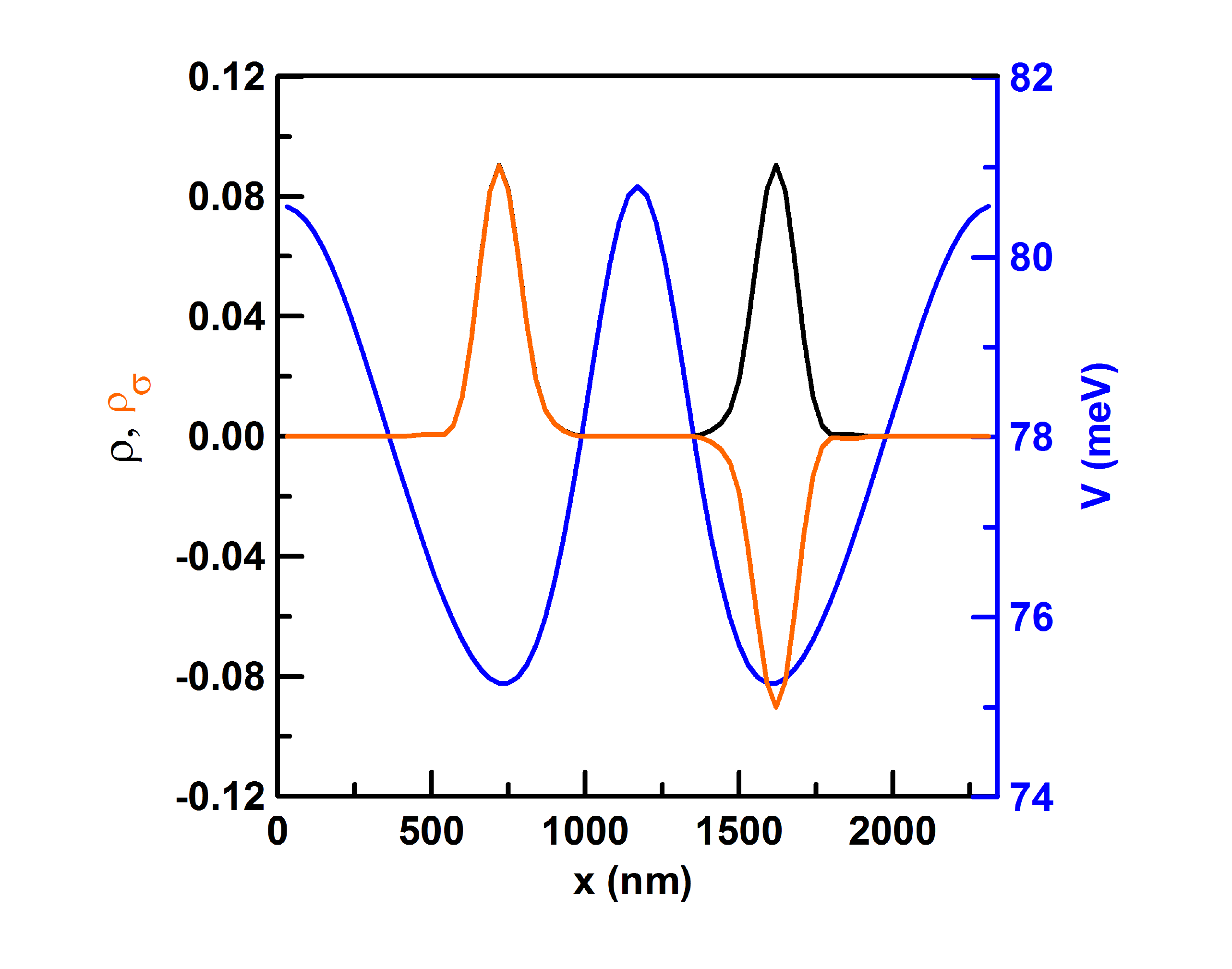}
\caption{\label{fig17}Final shapes of the spin density divided into two spatially separated parts with opposite spins (red curve), the total density (black curve), and the confinement potential at $z_0=L_z/2$  (blue curve). The densities are defined according to (Eqs. \ref{intsdens}, \ref{inttotdens}).}
\end{figure}

Oscillations, however, can be easily eliminated by subsequent elevation of the barrier at the moment when the velocity of spin density parts (i.e. $\frac{d}{dt}x_\uparrow(t)$ and $\frac{d}{dt}x_\downarrow(t)$) is zero. Fig.~\ref{fig15} shows a good candidate for such a moment at $t_2=127$~ps. The barrier is elevated by decreasing the voltage applied to the $\mathrm{U_0}$ gate by additional $100$~mV (now $U_0(t_2)=U(t_1)-260$~mV). Fig.~\ref{fig16} shows results of a simulation for that case,
with no rise  in the final part of the $\sigma_z^{\mathrm{R}}(t)$.
Fig.~\ref{fig17} shows the wavepacket and the potential barrier in the nanodevice center at the moment of maximal separation.
This time we compare the spin and total electron densities calculated along the $x$-axis but, unlike the 1D case, here also integrated along the $z$ direction. We use the following formulae:
\begin{equation}\label{intsdens}
\begin{split}
\rho_\sigma(x,t)&=\int_0^{L_z}\mathrm{d}z\,\Psi^\dag(x,z,t)\hat{\sigma}_z\Psi(x,z,t)\\
&=\int_0^{L_z}\mathrm{d}z\,\left(|\psi_{\uparrow}(x,z,t)|^2-|\psi_{\downarrow}(x,z,t)|^2\right),
\end{split}
\end{equation}
\begin{equation}\label{inttotdens}
\begin{split}
\rho(x,t)&=\int_0^{L_z}\mathrm{d}z\,\Psi^\dag(x,z,t)\Psi(x,z,t)\\
&=\int_0^{L_z}\mathrm{d}z\,\left(|\psi_{\uparrow}(x,z,t)|^2+|\psi_{\downarrow}(x,z,t)|^2\right),
\end{split}
\end{equation}
The potential energy has two minima with a barrier between them. Both parts of the spin density are separated and oscillate closely around these minima, but the amplitude of oscillation is much smaller than in the case in which the potential barrier was elevated only once. If we raise a barrier of full height at the first time by lowering the voltages by the full $250$~mV at time $t_1$, the amplitude of the oscillations will be significantly higher.

We mention that our proposal is robust against decoherence effects. 
The presented operation times on the electron spin are about $100$~ps for the complete electron spin density
separation---see e.g. Figs.~\ref{fig11} or \ref{fig14}, 
and are much shorter than the electron spin coherence times in InSb material.\cite{cohh}
These operation times might be further improved (reduced) by increasing the frequency $\omega/2\pi$.

Detection of the obtained Schr\"odinger's cat states could be based on measuring a fraction of charge of a definite spin, so called spin-to-charge conversion, using a quantum point contact (QPC) interface.

\section{Summary}
We proposed a design for a nanodevice based on a typical planar semiconductor heterostructure in which it is possible to create an entangled state of Schr\"odinger's cat type. This is achieved by spatial separation of the electron densities corresponding to opposite spin directions. Such a state has been generated in quantum optics \cite{winelandzrodlo, blatt, wineland1}. 

To create this state in our nanostructure, we need to use coherent states of the harmonic oscillator which are obtainable only for ideally parabolic confinement potentials. Generation of such a potential is one of the most important results of this paper. Nearly ideal parabolic potential along the $x$-axis is achievable in the proposed multi-gate nanodevice (see: Figs.~\ref{fig9} and \ref{fig10}) if we neglect the interaction of the electron confined in the quantum well with the induced charge on the device gates. This interaction causes self-focusing of the electron density, making spin separation much harder. A remedy for this effect is defocussing achieved by covering the entire nanostructure with a dielectric of lower permittivity compared to the well material. 
We have shown that self-focusing can be effectively compensated by adding a dielectric layer with simultaneous careful adjustment of the inner gate (rails) geometry.

We have performed simulations of numerous variants of our nanodevice via solution of the time-dependent Schr\"odinger equation with simultaneous tracking of the potential via solution of the Poisson's equation at every time step. As a result, we were able to take into account the geometry details, varying voltages applied to gates and changes of the electron density. Including also oscillating spin-orbit interaction, which induces the spin separation effect. 

\begin{acknowledgments}
This work has been supported by National Science Centre, Poland, under UMO-2014/13/B/ST3/04526. 
\end{acknowledgments}

\appendix*
\section{The Rashba spin-orbit coupling in semiconductor heterostructures}
Here we will briefly review valid models of the Rashba effect in III-V asymmetric semiconductor heterostructures.
We will make calculations of the spin-orbit coupling for these models, and compare them showing the validity of our chosen model (Eq.~\ref{raszka}).

Starting with the earliest works of Bychkov and Rashba\cite{bychkov,rashba2}, where they introduce discussed models, electrons confined at the heterojunction form two-dimensional electron gas (with $x$ and $z$ degrees of freedom):
\begin{equation}
H_{R}=\alpha\, \mathbf{k}^\mathrm{\scriptscriptstyle 2D}\!\times\hat{\mathbf{y}}\cdot\boldsymbol{\sigma},
\label{raszka2d}
\end{equation}
with two-dimensional electron momentum $\hbar\mathbf{k}^\mathrm{\scriptscriptstyle 2D}\!=\hbar(k_x,k_z)$, and versor $\hat{\mathbf{y}}$ normal to the junction surface.
The coupling strength $\alpha$ in the Hamiltonian (\ref{raszka2d}) is proportional to the external electric field $E_y$ (introducing asymmetry to the confinement) 
and some material dependent parameter $\alpha_\mathrm{so}$, which describes material details of the junction in $y$-direction: $\alpha=\alpha_\mathrm{so}[y]eE_y$. 
Moreover in [\onlinecite{winkler}] we find a summary of the famous discussion between Ando\cite{ando} and Lassnig\cite{lassnig}, which states that the Rashba spin-coupling in the conduction band results from the electric field in the valence band: $\langle E^{v}\rangle_c$.

In work [\onlinecite{starasilva}] the authors skip the band offset at the junction assuming an infinite barrier. Therefore, when calculating the $\alpha_\mathrm{so}$, they take the parameters of the conduction and valence bands only for the well material, using formula:
\begin{equation}
\alpha_\mathrm{so}=\frac{\hbar^2}{2m}\frac{\Delta}{\mathcal{E}_g}\frac{2\mathcal{E}_g+\Delta}{(\mathcal{E}_g+\Delta)(3\mathcal{E}_g+2\Delta)}. 
\label{asoc}
\end{equation}
For the InSb band parameters\cite{jap} $\mathcal{E}_g=235$~meV, $\Delta=810$~meV, $m=0.0135$~$m_e$ we get from above formula  $\alpha_\mathrm{so}=5.125$~$\mathrm{nm}^2$. 
A more complicated than (\ref{asoc}) but also a more accurate formula [\onlinecite{winkler}, Eq.~6.22] gives a similar value $5.23$~$\mathrm{nm}^2$.
The coupling is proportional to the external field $\mathbf{E}$, which in general is three-dimensional [\onlinecite{winkler}, Eq.~6.9]. 
In this work we assume a similar model, by replacing $\alpha\hat{\mathbf{y}}=\alpha_\mathrm{so}eE_y\hat{\mathbf{y}}\rightarrow\alpha_\mathrm{so}e\mathbf{E}$ in (Eq.~\ref{raszka2d}), thus obtaining (Eq.~\ref{raszka}), and taking the parameter value of $\alpha_\mathrm{so}=5.23$~$\mathrm{nm}^2$.

\begin{figure}
\vspace{2mm}
\includegraphics[width=0.47\textwidth]{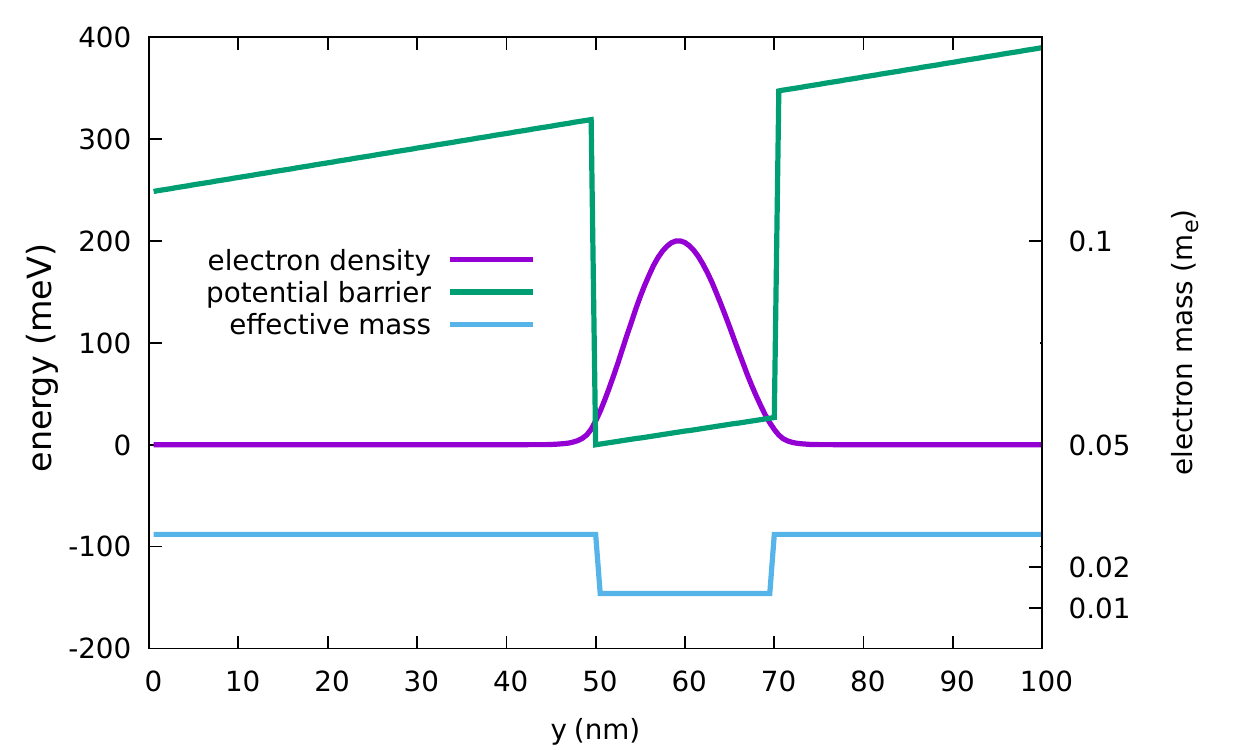}
\caption{\label{fig19}The conduction band profile (green curve) for the AlInSb/InSb/AlInSb heterostructure present in our device, together with the electron density (magenta) obtained for the Hamiltonian\cite{nowasilva} (Eq.~\ref{aham}) with the discontinuous effective mass (blue). Assumed electric field $E_y =1.33$~mV/nm. }
\end{figure}

Let us now see if our model\cite{starasilva,winkler} is sufficient 
and compare it with the model presented in [\onlinecite{nowasilva}]. The latter is a refinement of [\onlinecite{starasilva}] and is complete in sense, that it contains the confinement details within the heterostructure quantum well in the growth direction ($y$)---including different material parameters on both sides of the junction (effective masses, coupling values and bands parameters). Now we assume a single electron confined in the $y$ direction within our heterostructure, i.e. in a $20$~nm width  $\mathrm{Al_xIn_{1-x}Sb}$/InSb/$\mathrm{Al_xIn_{1-x}Sb}$ quantum well (with $x=25$~$\%$) and the conduction band offset $v_0=320$~meV together with the external field $E_y$. Such a conduction band profile for the heterostructure used in our device is shown in Fig.~\ref{fig19} (green curve).

Let us now assume the effective Hamiltonian in the $y$-direction for a conduction electron\cite{nowasilva} (nonparabolicity is negligible here) as:
\begin{equation}
H^c(y)=
  \begin{cases}
    -\frac{\hbar^2}{2m}\frac{d^2}{dy^2}+|e|E_y y     & \ y\text{ in InSb,}\\
    -\frac{\hbar^2}{2\mathcal{M}}\frac{d^2}{dy^2} + v_0 +|e|E_y y   & \ \text{elsewhere. }\\
  \end{cases}
  \label{aham}
\end{equation}
Note that the effective mass is discontinuous here, with the value of $m=0.0135$~$m_e$ for the InSb well, and $\mathcal{M}=0.028$~$m_e$ outside in the 
$\mathrm{Al_{\scriptstyle 0.25}In_{\scriptstyle 0.75}Sb}$ barriers region. The band mass in heterostructure is denoted by the blue curve in Fig.~\ref{fig19}.

For AlSb band mass is equal $m_\mathrm{\scriptscriptstyle AlSb}=0.140$,\cite{jap} thus  for the barrier alloy we assume $m_b=0.75\,m_\mathrm{\scriptscriptstyle (InSb)} +0.25\,m_\mathrm{\scriptscriptstyle AlSb}\simeq0.045$~$m_e$. 
Similarly, for the AlSb material we get $E^b_g=773$~meV, $\Delta^b=777$~meV and, resulting from (Eq.~\ref{asoc}), the spin orbit coupling $\alpha^b_\mathrm{so}=0.33$~$\mathrm{nm}^2$. The band mass and the spin-orbit coupling should be further renormalized in the barrier region. Using formulas [Eqs.~7 and 9 from \onlinecite{nowasilva}] we finally obtain $\mathcal{M}\simeq 0.62\,m^b=0.028$~$m_e$ and $\mathcal{A}_\mathrm{so}\simeq 3.35\,\alpha^b_\mathrm{so}=1.10$~$\mathrm{nm}^2$.

For these parameters we find eigenstates of the Hamiltonian (Eq.~\ref{aham}). We assume that the electron confined in the junction is in the ground state $\psi_0(y)$. For such a narrow quantum well ($20$~nm) the 1\textsuperscript{st} excited state is separated on the energy scale by about 120 meV. In a quantum well of height of 320 meV and width of 20 nm we have 3 bound states of energies: $39.82$~meV, $154.05$~meV and $287.25$~meV. For this calculated conduction electron density $\rho(y)=|\psi_0(y)|^2$ (magenta in Fig.~\ref{fig19}) we can finaly calculate the average value of the Rashba coupling in the heterostructure:
\begin{equation}
\alpha^h_\mathrm{so}=\int\limits_{y \text{ in InSb}}\!\!\!\mathrm{d}y \rho(y)\,\alpha_\mathrm{so} +\int\limits_{y\text{ in barrier}}\!\!\!\!\mathrm{d}y\rho(y)\,\mathcal{A}_\mathrm{so}.
  \label{avg}
\end{equation}
Note that the $\alpha^h_\mathrm{so}$ depends on the external electric field $E_y$ via the Hamiltonian (\ref{aham}).

\begin{figure}
\includegraphics[width=0.45\textwidth]{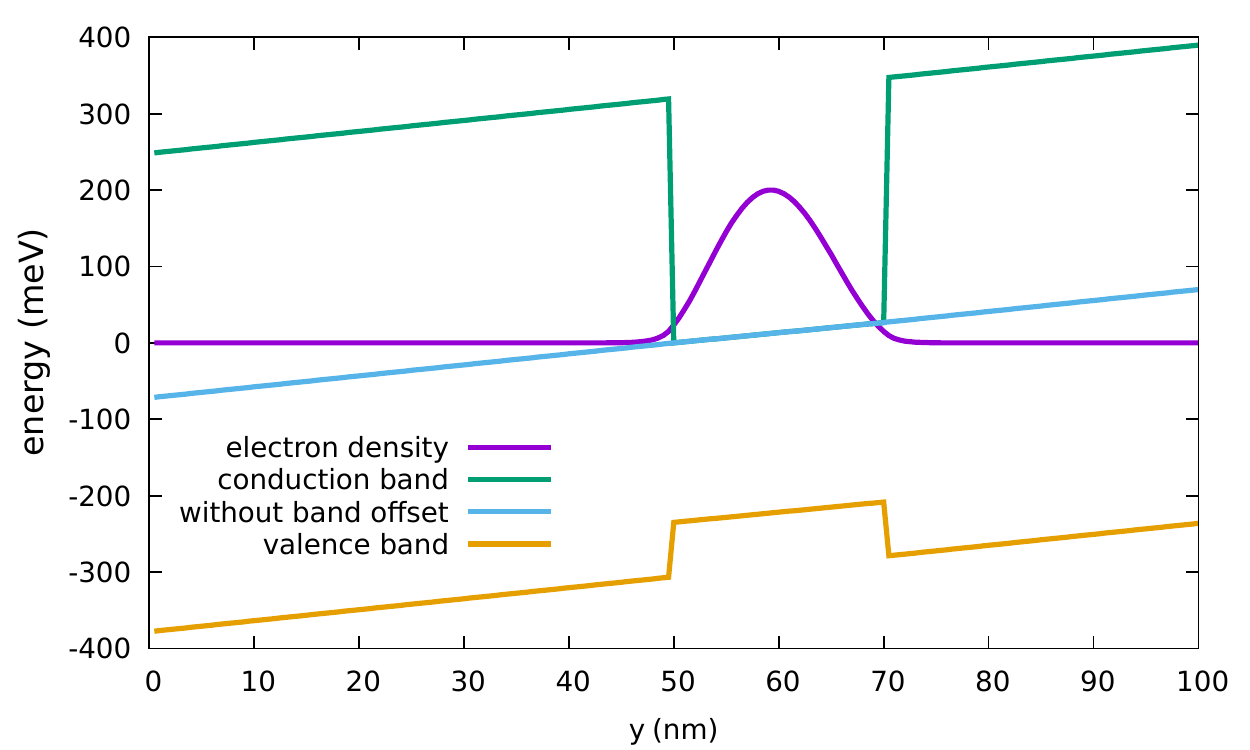}
\caption{\label{fig20}The conduction (green curve) and the valence band profile (orange one) for the heterostructure. In the simple toy model the electiv field from the valence band averaged with the conduction electron density (magenta) allows to determine the spin-orbit coupling.}
\end{figure}

For comparison, we additionally calculate the RSOI coupling value in our heterostructure within the simple toy model $\alpha^t_\mathrm{so}$.\cite{winkler} The $\alpha$ coupling is here simply calculated as product of the $\alpha_\mathrm{so}$ (in InSb) and the average electric field within the valence band $E^v(y)$ (orange curve in Fig.~\ref{fig20}) weighted with the electron density (magenta curve) in the conduction band (green curve): $\alpha_\mathrm{so}\langle E^{v}(y)\rangle_c$. 
For a given $E_y$ the effective coupling equals $\alpha^t_\mathrm{so}=\alpha_\mathrm{so}\langle E^{v}\rangle_c/E_y$. The electron density is calculated here as in Fig.~\ref{fig19}. We compare this result with the analytical expression for $\langle E^{v}\rangle_c=\frac{v_0+v_v}{v_0}E_y$ [Eq.~6.33 from \onlinecite{winkler}], 
with the offset in the valence band $v_v/v_0\simeq0.22$,\cite{jap} giving $\alpha^a_\mathrm{so}\simeq1.22\,\alpha_\mathrm{so}$.

\begin{figure}[b]
\includegraphics[width=0.45\textwidth]{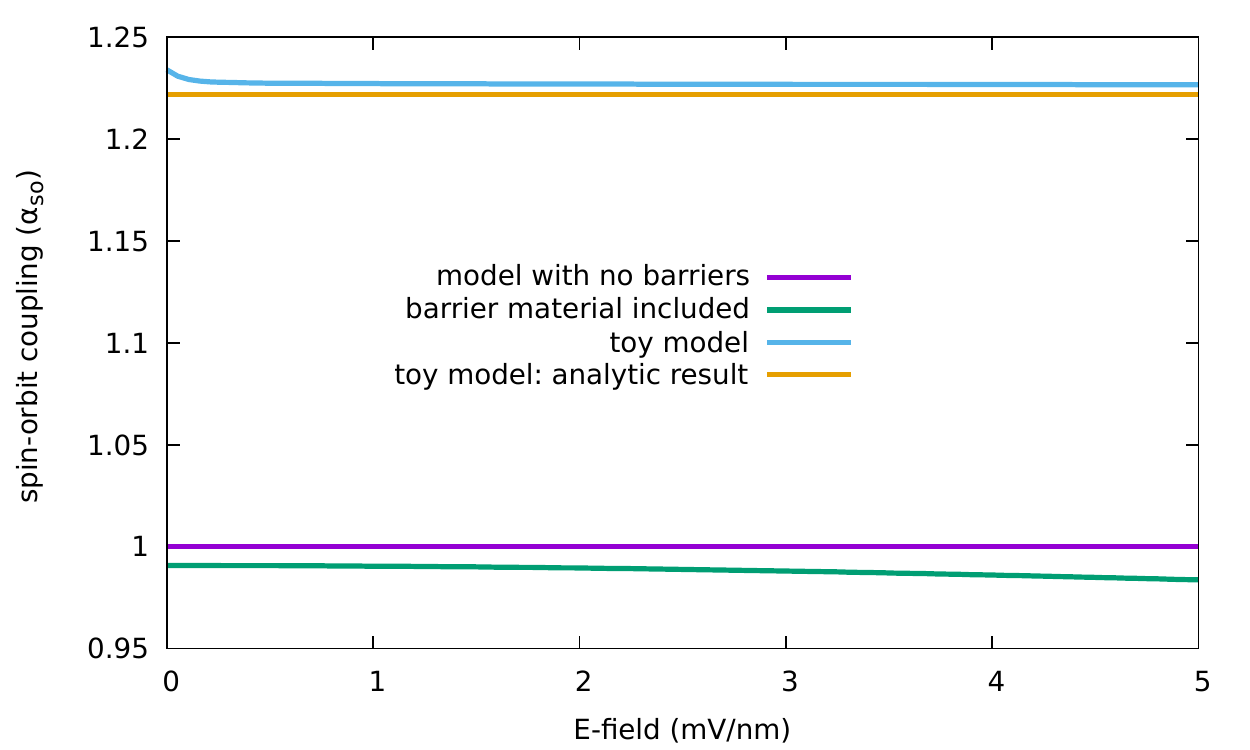}
\caption{\label{fig21} Comparison of the all presented models. The toy models (blue and orange curves) overestimate the RSOI coupling, while the most accurate model (green curve) gives only small (about 1\%) correction to the one assumed in the work (magenta curve).}
\end{figure}

Let us now compare all the presented models as a function of the external electric field $E_y$ assuming values, as in our nanodevice, between $0.33$ and $2.33$ mV/nm.
In Fig.~\ref{fig21} we plotted $\alpha^h_\mathrm{so}$ for: the most accurate model \cite{nowasilva}, taking into account confinement details in $y$ direction as well as the barrier material (green curve); $\alpha^t_\mathrm{so}$ the simple toy model calculating a mean field in the valence band \cite{winkler} (blue curve). Both are compared with $\alpha_\mathrm{so}$ from a model that neglects details about the barrier in the junction\cite{winkler, starasilva} (magenta) and the analytic solution from a toy model\cite{winkler} $\alpha^t_\mathrm{so}$. 
Last two curves are independent of the electric field $E_y$. Values for the toy model (blue and orange) slightly overestimate the value we assumed (magenta). Nevertheless the most accurate and complex model (green) produces results that are about 1\% lower for the range of electric fields used in our device (not greater than about $2.33$~$\mathrm{mV}\!/\mathrm{nm}$). 

According to presented calculations we conclude that within the range of electric fields present in our device, corrections to $\alpha^h_\mathrm{so}$ taking into account details of confinement in the junction are negligible (of the order of 1\%). Thus in our calculations we use a simpler spin-orbit coupling model, which takes into account only the QW material, for which we get the coupling constant $\alpha_\mathrm{so}=5.23$~$\mathrm{nm}^2$.

\bibliography{thebibliography}

\end{document}